\documentclass[twocolumn,showpacs,%
aps,superscriptaddress,
prd,notitlepage,showkeys,
nofootinbib]{revtex4-1}

\usepackage{ulem}
\usepackage{amssymb}
\usepackage{amsmath}
\usepackage{graphicx}
\usepackage{dcolumn}
\usepackage[colorlinks,urlcolor=blue,citecolor=blue,linkcolor=blue]{hyperref}
\usepackage{color,units}
\usepackage[dvipsnames]{xcolor} 
\usepackage{lineno}
\usepackage{xspace}
\usepackage{longtable} 
\usepackage{float} 
\usepackage{multirow}
\usepackage{comment}
\usepackage{braket}
\usepackage{hyperref}
\usepackage{amsfonts,wasysym,epsfig,verbatim,subfigure,bm,mathrsfs,lipsum}
\begin{document}
\newcommand{\IUCAA}{Inter-University Centre for Astronomy and
Astrophysics, Post Bag 4, Ganeshkhind, Pune 411 007, India}

\newcommand{\MPI}{Max-Planck-Institut f{\"u}r Gravitationsphysik (Albert-Einstein-Institut), D-30167 Hannover, Germany}

\newcommand{\LBNZ}{Leibniz Universit{\"a}t Hannover, D-30167 Hannover, Germany}

\title{Probing horizon scale quantum effects with Love}
\author{Sayak Datta}\email{sayak.datta@aei.mpg.de} 
\affiliation{\IUCAA}\affiliation{\MPI}\affiliation{\LBNZ}
\date{\today}

\begin{abstract}
Future gravitational wave detectors have been projected to be able to probe the nature of compact objects in great detail. In this work, we study the potential observability of the small-length scale physics near the black hole horizon with the tidal deformability of the compact objects in an inspiraling binary. We find that it is possible to probe them with extreme mass ratio inspirals. We discuss how the quantum effects can affect gravitational wave observables. This as a consequence is bound to shape our understanding of the quantum scale near the horizon.
\end{abstract}

\maketitle

\section{Introduction} The discovery of gravitational waves(GWs) \cite{Abbott:2016blz, LIGOScientific:2018mvr} paved the way towards probing fundamental physics. These observations provided a fillip to tests of General Relativity (GR) in the strong-field regime~\cite{LIGOScientific:2019fpa,Abbott:2018lct};
e.g., stringent bounds on the mass of the graviton and violations of Lorentz invariance have been placed~\cite{TheLIGOScientific:2016pea, TheLIGOScientific:2016src, Abbott:2017vtc}. As a result, GWs have become very important in the context of fundamental physics. Various possible distinctions between black holes (BHs) and other exotic compact objects (ECOs) based on tidal deformability \cite{Cardoso:2017cfl, Sennett:2017etc, Maselli:2017cmm, Brustein:2020tpg}, tidal heating \cite{Maselli:2017cmm, Datta:2019euh, Datta:2019epe, Datta:2020rvo, Agullo:2020hxe, Chakraborty:2021gdf, Sherf:2021ppp, Datta:2021row, Sago:2021iku, Maggio:2021uge, Sago:2022bbj, Mukherjee:2022wws}, multipole moments \cite{Krishnendu:2017shb,Datta:2019euh, Bianchi:2020bxa, Mukherjee:2020how, Datta:2020axm, Narikawa:2021pak}, {\it echoes} in postmerger \cite{Cardoso:2016rao, Cardoso:2016oxy, Maggio:2019zyv, Tsang:2019zra,Abedi:2016hgu, Westerweck:2017hus, Cardoso:2019rvt, Chen:2020htz, Xin:2021zir} and electromagnetic observations \cite{Titarchuk:2005rr,Bambi:2013sha,Jiang:2014loa,Bambi:2015kza,Bambi:2015kza, Cardoso:2019rvt} has been proposed in the literature.

One of the very intriguing questions in fundamental physics is how gravity behaves in the quantum regime. Since GWs bring information from the very close vicinity of BHs, it is expected that GWs may shed some light on this mystery \cite{Jenkins:2018ysa, Calmet:2018rkj, Cardoso:2019apo, Laghi:2020rgl, Agullo:2020hxe, Datta:2021row, Chakraborty:2021gdf}. The idea behind such expectations follows from the fact that the Planck scale physics may affect the tidal Love numbers (TLNs) of the compact objects \cite{Maselli:2017cmm, Brustein:2020tpg, Brustein:2021bnw, Bonelli:2021uvf}. As compact objects coalesce, the information of the TLNs gets imprinted on the emitted GWs.

We study the challenges in achieving this due to the statistical error and the quantum noise. We will demonstrate for the first time that despite the quantum noise, it is possible to probe the near-horizon quantum scale physics with extreme mass ratio inspirals. As a result, not only do the small quantum corrections to the values of TLNs become measurable, but also inferring quantum noise will be possible. This will inevitably bring information from the quantum world near the horizon, shaping our understanding of the quantum nature of gravity.

In Sec. \ref{Sec:TD} we will discuss the basics of tidal deformability. Then in Sec. \ref{Sec: delta-k} the $\delta-k$ relation will be discussed. In Sec. \ref{Sec:Q-noise} the impact 
of quantum noise will be investigated. In \ref{Sec:Observability}  we will investigate the observability of small Love numbers with LISA. In Sec. \ref{Sec:Invalidity} limitation of the $\delta-k$ relation will be discussed. Then in Sec. \ref{subsec:validity} a formalism will be constructed that is applicable for computing quantum contribution to the Love numbers. Finally, in Sec. \ref{Sec:discussion} we will discuss the implication of our work and also its limitations.

\section{Tidal deformability} 
\label{Sec:TD}

Consider a binary with the mass of the $i$th component to be $m_i$ in the inspiral phase. We can model these systems using the post-Newtonian (PN) theory, which is a weak-field/slow-velocity expansion of
the field equations. The emitted GWs from such systems can be modeled in the frequency domain as \cite{Maselli:2017cmm, Datta:2020gem},
\begin{equation}
\Tilde{h}(f) = A(f)e^{i(\psi_{PP}(f)+\psi_{TH}(f)+\psi_{TD}(f))}, 
\end{equation}
where $f$ is the GW frequency, $A(f)$ is the 
amplitude in the frequency domain. $\psi_{PP}(f)$ is the contribution to the GW Fourier phase while treating the objects as spinning point particles, $\psi_{TH}(f)$ is the contribution due to tidal heating, and $\psi_{TD}(f)$ is the contribution due to their tidal deformability. In several works, it has been argued that $\psi_{TH}(f)$ and $\psi_{TD}(f)$ can be used to probe the nature of compact objects. As a result, it can be used as a distinguisher between black holes (BHs) and exotic compact objects (ECOs). In this work, we will focus only on $\psi_{TD}(f)$. To leading PN order, this contribution for circular equatorial orbits is \cite{Flanagan:2007ix}
\begin{equation}
\psi_{TD}(f) = -\frac{117}{8}\frac{(1+q)^2}{q}\frac{\tilde{\Lambda}}{m^5}v^5,
\end{equation}
where $v = (\pi mf)^{1/3}$ is the velocity, with $m = m_1 + m_2$ the total mass, and 

\begin{equation}
26\tilde{\Lambda} = (1 + 12/q) \lambda_1 + (1 + 12q)\lambda_2,
\end{equation}
where, $\lambda_i = \frac{2}{3} {\textsl k}_i m_i^5$ with ${\textsl k}_i$ the ($\ell= 2$, electric-type) TLNs and $q = m_1/m_2$ is the mass ratio.

\section{$\delta - {\textsl k}$ relation}
\label{Sec: delta-k}

TLNs are the response of a body to an external tidal field. It explicitly depends on the details of the internal structure of the compact object. It has been argued that for the BHs of GR, the TLN vanishes \cite{Binnington:2009bb, Landry:2014jka,LeTiec:2020bos, Chia:2020yla}\footnote{Recently in several works it has been demonstrated that the origin of the vanishing Love number is connected with the so-called Ladder symmetry \cite{Charalambous:2021kcz, Charalambous:2021mea, Hui:2021vcv, Hui:2022vbh, BenAchour:2022uqo}. In the presence of quantum hair this may break down.}. Other compact objects unlike BHs, have a non-zero TLN. According to their equation of state, matter anisotropy, and fluid nature, neutron stars can have TLNs of $\mathcal{O}(10^2)$ \cite{Hinderer:2007mb, Hinderer:2009ca, TheLIGOScientific:2017qsa, Abbott:2018exr, Char:2018grw, Datta:2019ueq, Raposo:2018rjn, Biswas:2019gkw, Baiotti:2019sew, Dietrich:2020eud} and similarly for the boson stars \cite{Sennett:2017etc, Cardoso:2017cfl}. TLNs of some highly compact ECOs scales as $\sim 1/| \log(\epsilon)|$, where $\delta \equiv r_s - r_H \equiv \epsilon r_H$, where $r_s$ is the actual surface position of the ECO, and $r_H$ is the surface position of the horizon if it were a BH \cite{Cardoso:2017cfl}. 

Motivated by this finding, it was argued in Ref. \cite{Maselli:2017cmm} that this logarithmic behavior can be used to possibly probe the Planck scale physics near the horizon (surface) of a BH (ECO). This logarithmic behaviour translates to the $\delta-{\textsl k}$ relation as follows (caveats are discussed later) \cite{Maselli:2017cmm},
\begin{equation}
\label{eq:delta-k}
\delta = r_s - r_H = r_He^{-1/{\textsl k}}
\end{equation}

Deviation of Planckian order ($\delta=\ell_{pl}\sim\mathcal{O}(10^{-35})${\rm meters}) corresponds to ${\textsl k} \sim 10^{-2}$ for masses of the BH ranging in the range $(10^5-10^7)M_{\odot}$ \cite{Maselli:2017cmm, Addazi:2018uhd}. \footnote{For an invariant definition of $\delta$ check \cite{Addazi:2018uhd}.} From this it was proposed that by measuring small $k$, Planck scale physics can be probed.

\section{Measuring quantum noise}
\label{Sec:Q-noise}

In such a case, it would seem that the only limitation disallowing us from such achievement is the sensitivity of the detectors. However, in \cite{Addazi:2018uhd} it has been argued that it is unlikely to be the case, as quantum noise of $\delta$ will populate at that level. As a result, the error in $\delta$ will get modified as \cite{Addazi:2018uhd},

\begin{equation}
\label{systematic1}
\begin{split}
\frac{\sigma_{\delta}^{\rm Tot}}{\bar{\delta}} =&  \sqrt{\left(\frac{\sigma_{r_H}^{\rm Stat}}{\bar{r_H}}\right)^2 + \frac{1}{\bar{{\textsl k}}^2}\left(\frac{\sigma^{Stat}_{\textsl k}}{\bar{{\textsl k}}}\right)^2 + \frac{a^2\ell_{pl}^2}{\bar{\delta}^2}}\\
\equiv& \sqrt{\left(\frac{\sigma_{\delta}^{\rm Stat}}{\bar{\delta}}\right)^2 + \left(\frac{\sigma_{\delta}^{\rm Sys}}{\bar{\delta}}\right)^2}
\end{split}
\end{equation}
where, $\bar{\delta}$ and $\bar{r}_H$ is the estimated value of $\delta$ and $r_H$ from the observation, and $\sigma_{\delta}^{\rm Sys} = a^2\ell_{pl}^2$. $\sigma^{\rm Stat}_{r_H}$, $\sigma^{\rm Stat}_{k}$, $\sigma^{\rm Stat}_{\delta}$ are the statistical error in $r_H$, $k$ and $\delta$ respectively. ${\rm Stat}$ is the shorthand for statistical error. The error induced by quantum noise is $a\ell_{pl}$. Where $\ell_{pl}$ is the Planck length and $a$ is a number $\sim \mathcal{O}(1)$.

Assuming this behavior of error, we can estimate $a$, which will help us in measuring the quantum noise. It is the first key observation of the current work. This will be possible to do since other parameters can be measured independently. From the observation we will have $\sigma^{\rm Stat}_{M}$, $\sigma^{\rm Stat}_{\chi}$, $\bar{M}$, $\bar{\chi}$. This can be used to estimate $\sigma^{\rm Stat}_{r_H}$, $\bar{r}_H$. From the observation, the inferred value of TLN $\bar{{\textsl k}}$ will also be available. Therefore, if we can have an estimation of $\sigma_{{\textsl k}}^{\rm Stat}$ then we can estimate the $a^2$. 

This can be done by performing simulations with injected synthetic signal in detectors with $\bar{{\textsl k}}$ and other observed parameters. Running a Bayesian estimation on that we can have an estimation of the statistical error, which is an artifact of the observation. With sufficiently sensitive detector $\sigma_{{\textsl k}}^{\rm Stat}$ can be reduced to very small values. By estimating those values from simulations we can estimate the systematic error, which is arising from the quantum nature. Having an estimation of $a^2$ can lead us to understand the quantum states near the horizon. For this purpose, in the next section, we will investigate if it is possible to reduce the statistical error sufficiently in future detectors. 

\section{Observability}
\label{Sec:Observability}

Extreme mass ratio inspirals (EMRIs) are one of the promising sources of GW which will be observed with the future space-based Laser Interferometer Space Antenna (LISA) \cite{Audley:2017drz}. The emitted GW from these systems can stay in the detector band from months to a year. As a result, despite being small, with LISA we will be able to measure the TLNs of supermassive BHs in EMRI, quite precisely. Although the rates of EMRIs are not well understood it is expected that several such sources will be detected with LISA \cite{Gair:2004iv, Gair:2012vi, Amaro-Seoane:2019umn}.

\begin{figure}
\centering
\includegraphics[width=\linewidth]{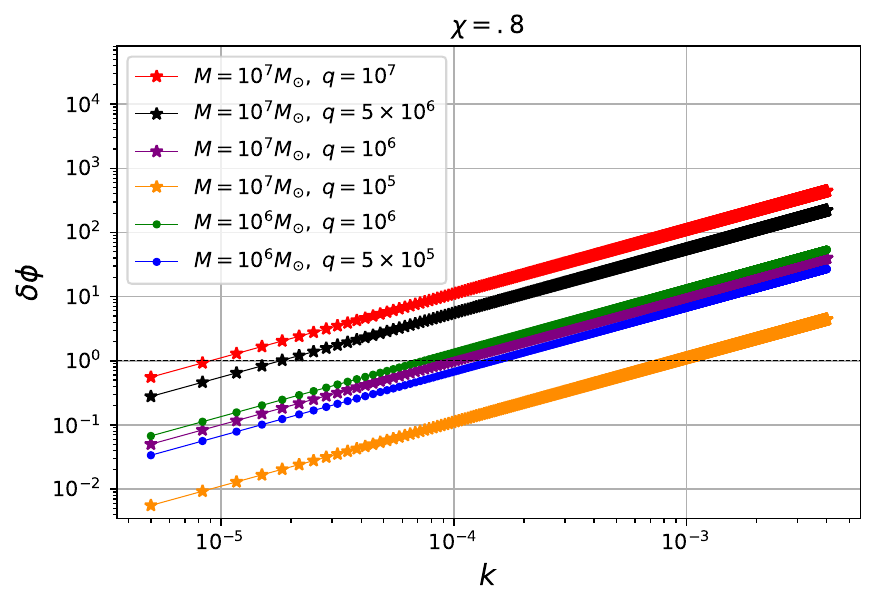}
\caption{We show the magnitude of dephasing $(\delta \phi)$ in radian, as a function of ${\textsl k}$. We varied $M$ and $q$ while keeping $\chi=.8$.}
\label{fig:EMRI_phase}
\end{figure}

\begin{figure}
\centering
\includegraphics[width=\linewidth]{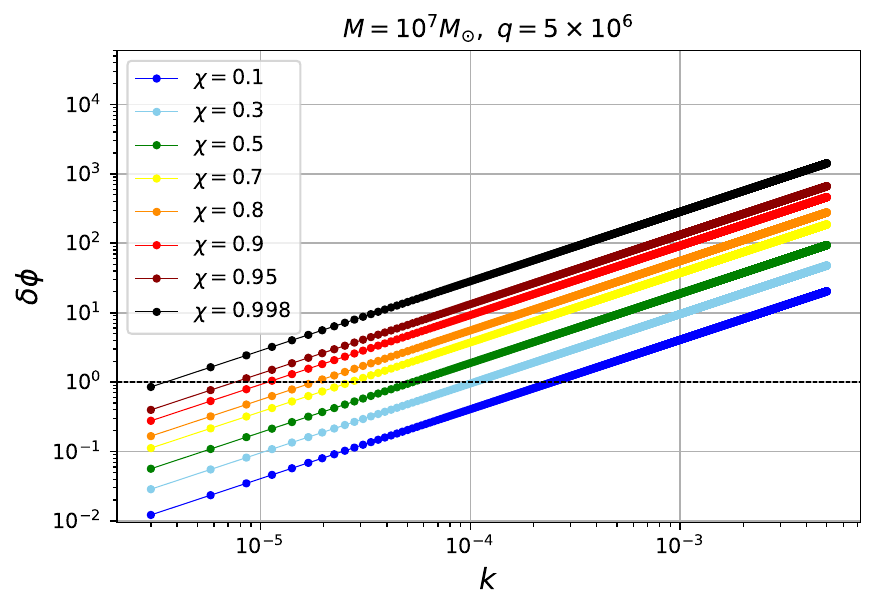}
\caption{We show the magnitude of the dephasing $(\delta \phi)$ in radian, as a function of ${\textsl k}$.  We varied $\chi$ while keeping primary and secondary mass fixed at $10^7M_{\odot}$ and $2M_{\odot}$ respectively.}
\label{fig:EMRI_phase_spin}
\end{figure}

\begin{figure}
\centering
\includegraphics[width=\linewidth]{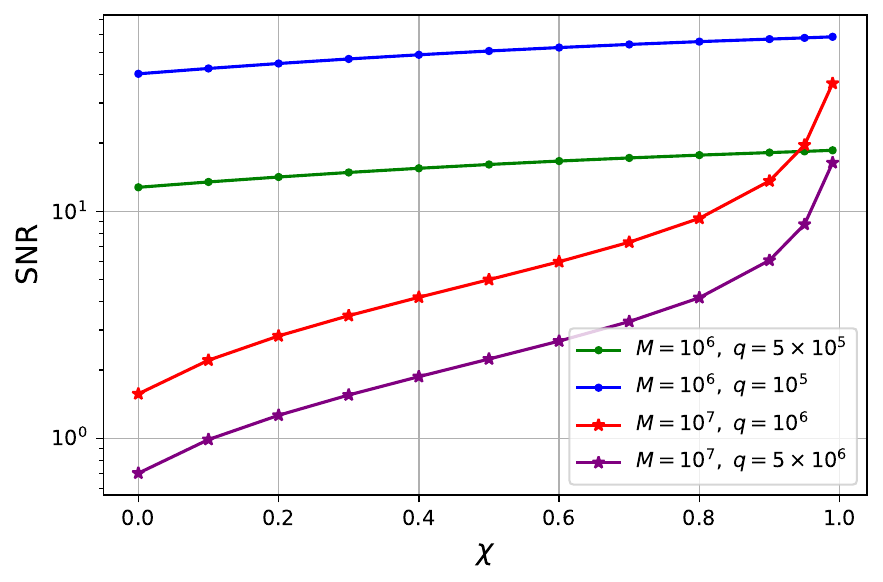}
\caption{In the above figure we demonstrate the SNR. The signal from $.4{\rm mHz}$ to ISCO frequency is considered. The SNR is lesser for total mass $10^7 M_{\odot}$ compared to $10^6 M_{\odot}$. This is both due to a higher mass ratio and shorter duration of signal in the observable band. The sources are considered to be at $1 {\rm GPc}$. }
\label{fig:EMRI_SNR}
\end{figure}

To estimate the effect of the TLN of these supermassive bodies in EMRI, we calculate dephasing as a function of ${\textsl k}$. We ignore the contribution of the secondary body. The primary body's mass is considered to be $m_1=M$ and the dimensionless spin is $\chi$. A useful estimator to describe the effects of ${\textsl k}$ in the phase is the total number of GW cycles $(\equiv N)$ that accumulates within a given frequency band of the detectors. In terms of the frequency-domain phase $\psi_{TD}(f)$ it is expressed as,

\begin{equation}
\label{eq:N}
\delta\phi = 2\pi N \equiv \int_{.4\,{\rm mHz}}^{f_{\rm ISCO}(M,\chi)} f df \frac{d^2\psi_{TD}(f)}{df^2},
\end{equation} 
where $f_{\rm ISCO}$ is the GW frequency at the innermost stable circular orbit. In Fig. \ref{fig:EMRI_phase} and Fig. \ref{fig:EMRI_phase_spin} we show the magnitude of the dephasing $(\delta \phi)$ in radian, as a function of ${\textsl k}$. The results are consistent with the expectations discussed in Ref. \cite{Pani:2019cyc}. The black dashed horizontal line represents $\delta \phi = 1$ radian. Dephasing $\delta\phi >1$ represents a strong effect \cite{Flanagan:1997kp,Lindblom:2008cm,Amaro-Seoane:2007osp, Porter:2010mb,Amaro-Seoane:2011fgy, Katz:2021yft}. In reality $\delta\phi >1/\rho $ is a much more pertinent condition for an effect to be detectable, where $\rho$ is signal to noise ratio (SNR) of the signal\footnote{For further details on the connection between dephasing, mismatch, and SNR, check appendix \ref{mismatch}.}. For this purpose, in Fig. \ref{fig:EMRI_SNR} we plot the SNR of several sources situated at 1GPc, computed between $.4$ mHz and ISCO frequency. To compute the SNR the considered LISA sensitivity curve has been taken from \cite{Robson:2018ifk}. As can be seen the SNR $\gtrsim 1$. If these same sources are nearby then the SNR will increase. For the sources considered the lowest value of SNR is $\sim 1$. Therefore for such sources required dephasing would be $\sim 1$. From the dephasing plot, it can be seen that it is possible to achieve. Note, $k\sim 10^{-2}$ corresponds to Planck scale, assuming the $\delta-k$ relation in Eq. \ref{eq:delta-k}. Therefore, the smaller values correspond to the sub-Planckian scale that should be dominated by Planck scale noise. The result implies that the EMRIs can be the potential sources that will be sensitive to small-scale physics. However, considering $\delta\phi >1$ as an observational threshold has limitations.  Although for very high SNR, this threshold can act as a sufficient condition to be detectable, it might not be good enough for low SNR sources. In such case, the statistical uncertainty on the phase could eventually overreach $\delta \phi$, if $k$ values are very small.

To be sensitive to the Planck scale physics
it is atleast necessary that $\sigma^{Stat}_{\delta}<\bar{\delta}$, where $\sigma^{Stat}_{\delta}$ is the statistical error of $\delta$. In EMRIs the statistical error in $\delta$ will be dominated by the statistical error in ${\textsl k}$ since a fractional error on mass and spin will be very less in LISA \cite{Poisson:1996tc, Barack:2003fp}. Hence,

\begin{equation}
\frac{\sigma^{Stat}_{\delta}}{\bar{\delta}} \sim \frac{\sigma^{Stat}_{\textsl k}}{\bar{{\textsl k}}^2}.
\end{equation}

Assuming Eq. (\ref{eq:delta-k}) to be valid, for $\bar{{\textsl k}}\sim .005\,(.01)$ to probe subPlanckian effects it is required that $\sigma^{Stat}_{\textsl k} < 2.5 \times 10^{-5}\,(10^{-4})$. From Fig. \ref{fig:EMRI_phase} and Fig. \ref{fig:EMRI_phase_spin} it can be observed that such sensitivity can be reached with EMRIs. Hence, statistical error is low enough in EMRIs. This does not mean that the Planck scale physics can be probed with this accuracy. It means that the dominating error will be just the quantum noise described in Eq.(\ref{systematic1}). As discussed before it can be used to estimate the quantum error itself, assuming the $\delta-k$ relation to be true. But in later sections, we will discuss why it is not just to assume the $\delta-k$ relation apriori. Rather we should use this opportunity to do an accurate measurement of the $k$ to probe quantum correction or alternate theories of gravity. As well as we should try to investigate if there is any quantum error associated with $k$. The measurement of quantum error in $k$ does not require $\delta-k$ relation to be valid apriori as it can arise from near horizon quantum effects.
Note, the primary difference between the current work and Ref. \cite{Addazi:2018uhd} is that the considered sources are different. In the present work, the considered sources are EMRIs whereas, the sources considered in Ref. \cite{Addazi:2018uhd} and Ref. \cite{Maselli:2017cmm} are supermassive binaries.

\section{Invalidity of $\delta-{\textsl k}$ relation}
\label{Sec:Invalidity}

In this section, we will argue that Eq. (\ref{eq:delta-k}) is unlikely to hold in the context of GW observation. It is not justified to assume that ${\textsl k}\rightarrow 1/| \log(\epsilon)|$ scaling will be valid on a very small scale where quantum effects become important. This result has been derived assuming classical gravity. To probe small-scale physics, it is necessary for $\epsilon$ to be of that order. The conventional matter should collapse if it is distributed in such close proximity. The origin of such values of $\epsilon$ must be therefore exotic matter or quantum effects.

Hence, these systems are not ``classical" systems to begin with. Consequently, it will become necessary to take into account the quantum properties of the states of the system to find the sub-leading contribution to the leading order classical results. This sub-leading ``quantum corrections" most likely will be the interaction between the quantum observables at the quantum scale and the classical fields (discussed later). In such a case, the $\delta-{\textsl k}$ relationship is likely to get modified by ${\textsl k} \sim 1/| \log(\epsilon)|^n +{\textsl k}_q(\epsilon)$, with $n$ being a real number \cite{Brustein:2020tpg}. Therefore even though the first term starts to go to zero for very small $\epsilon$, the second term survives and captures the details of the quantum nature. For BH as ${\textsl k}= 0$ classically, quantum effects can introduce nonzero ${\textsl k}_q$, resulting in ${\textsl k} = {\textsl k}_q(\epsilon)$.

It is important to ask, from which value of $\epsilon=\epsilon_q$ this behavior becomes important. If the compact objects are not sufficiently compact i.e. $\epsilon_{ECO}\gg \epsilon_q$, then these quantum corrections ($k_q$) will not be important. However if $\epsilon_{ECO}\lesssim \epsilon_q$, they can be used to probe the quantum scale near the horizon that is larger than the Planck scale. Another key issue is if any kind of $\delta-k$ relation seizes to exist then relations like Eq.(\ref{systematic1}) become invalid, making $\delta$ immeasurable from the measurement of $k$. But there will exist $k_q$ and non-zero systematic quantum noise in $k$, which will be discussed in the next section. Therefore, precise measurement of $k$ and its error can help us probe the quantum nature near the horizon scale. As has already been demonstrated, EMRIs has such potential.

\section{Love in the Quantum world}\label{subsec:validity}

Due to the presence of an external tidal field, a nonzero quadrupole moment ${\bf Q}$ (multipole moment) gets induced on the bodies. In a linear regime, it is proportional to the external tidal field ${\bf \mathcal{E}}$, where the proportionality constant is the TLN $(k)$. In the $\delta \gg \ell_{pl}$ limit, a semiclassical quantum gravity approach can be applied to find the corrections to the classical contribution to the $k$.

Throughout our calculations, we will suppress the indices, and any non-scalar tensor will be represented by boldface. Therefore the tidal deformability can be defined as,

\begin{equation}
\label{eq:defromability}
\bf Q = -\lambda \mathcal{E}
\end{equation}
where, $\lambda =\frac{2}{3}{\textsl k} m^5$, with {\textsl k} and $m$ being the TLN and the mass of the body (note $m$ is not the total mass of a binary as was assumed before).

To find the contribution of the quantum effects we will consider quantum operators for all physical observables. We will assume none of the operators have zero eigenvalues, hence they are invertible \footnote{In reality this stringent condition may not be required as all the required operators are scaled by a classical value.}. We will separate the classical contribution and quantum fluctuation as,

\begin{equation}
\begin{split}
\lambda \rightarrow &\Hat{\lambda} + \lambda_c \hat{I}\\
\bf Q \rightarrow & \Hat{\bf Q} + Q_c \bf{\hat{I}}\\
\bf \mathcal{E} \rightarrow & \Hat{\bf \mathcal{E}} + \mathcal{E}_c \bf{\hat{I}}
\end{split}
\end{equation}
where $\lambda_c,\,{\bf Q_c},\,{\bf \mathcal{E}_c}$ are the classical contribution to the observables, and ${\bf \hat I}$ (${ \hat I}$) is the tensor (scalar) identity operator. We will also assume that Eq. (\ref{eq:defromability}) is valid in this regime but in the sense of quantum operators\footnote{It is likely that there will be some modification due to quantum effects. But for the current work, we will ignore such contributions.}. Hence, it can be expressed as,

\begin{equation}
{ \bf \Hat{Q}} + Q_c {\bf \hat I} = -\lambda_c \mathcal{E}_c {\bf {\hat I}- \Hat{\mathcal{E}} } \lambda_c - {\bf \Hat{\lambda}} \mathcal{E}_c - {\bf \Hat{\lambda} \Hat{\mathcal{E}}}.
\end{equation}
Using this relation it is possible to identify the expressions of the classical contributions as well as the quantum contributions as,

\begin{equation}
\lambda_c = -\frac{Q_c}{\mathcal{E}_c}, \,\, {\bf \Hat{\lambda}} =-\left(\frac{{ \Hat{\mathcal{E}}\lambda_c + \bf \Hat{Q}} }{\mathcal{E}_c + \Hat{\mathcal{E}}}\right) \approx - \frac{{ \bf \Hat{Q}}}{\mathcal{E}_c} + \mathcal{O}(\Hat{\mathcal{E}}).
\end{equation}
Note, ${\bf \hat{\mathcal{E}}}$ represents quantum corrections to the classical value of the external tidal field. Hence, this quantum correction represents the quantum correction of the external body's mass and the separation. In the right-most equation contribution of $\hat{\mathcal{E}}$ has been ignored.

This result is equivalent to the expressions used in Ref. \cite{Brustein:2020tpg, Brustein:2021bnw} (Check \cite{Bonelli:2021uvf, Kim:2020dif}). We will assume that the state of the system is $|\Psi\rangle$ and we will suppress the $\Psi$ while writing the expectation value with respect to $|\Psi\rangle$. As a result, deformability gets modified as,
\begin{equation}
\lambda = \lambda_c + \langle {\bf \hat{\lambda}}\rangle \equiv \lambda_c + \lambda_q,
\end{equation}
where $\langle\rangle$ represents expectation value, and,
\begin{equation}
\lambda_q = -\left\langle\frac{{ \Hat{\mathcal{E}}\lambda_c + \bf \Hat{Q}} }{\mathcal{E}_c + \Hat{\mathcal{E}}}\right\rangle \approx -\frac{\langle{\bf \Hat{Q}}\rangle}{\mathcal{E}_c}.
\end{equation}
Using this expression the systematic error in $\lambda$ arising from the quantum nature can be expressed as,

\begin{equation}
\sigma^{\rm Sys}_{\lambda} = \sqrt{\left\langle\left(\frac{{ \Hat{\mathcal{E}}\lambda_c + \bf \Hat{Q}} }{\mathcal{E}_c + \Hat{\mathcal{E}}} + \lambda_q\right)^2\right\rangle}
\end{equation}

As the statistical error in ${\textsl k}$ for EMRIs will be lower, observability of quantum noise solely will depend on the value of the standard deviation of the fluctuation of $\hat{{\textsl k}}$. To find the corresponding result in ${\textsl k}$, we separate out each observable into its classical and quantum parts as,
\begin{equation}
\begin{split}
{\textsl k} \rightarrow &\hat{{\textsl k}} + \hat{I}{\textsl k}_c, \,\,\,
m \rightarrow \hat{m} + \hat{I}m_c
\end{split}
\end{equation}
Using $\lambda=\frac{2}{3}{\textsl k} m^5$ we find,

\begin{equation}
\begin{split}
\lambda_c =& \frac{2}{3}m_c^5{\textsl k}_c,\\
\hat{{\textsl k}} =& \left(\frac{3\hat{\lambda}}{2m_c^5} - \frac{5{\textsl k}_c\hat{m}}{m_c} -\frac{15\hat{\lambda}\hat{m}}{2m_c^6}\right)  + \mathcal{O}(\hat{m}^2)\\
\lambda_q =& \frac{2}{3}m_c^5 \left({\textsl k}_q + 5{\textsl k}_c\frac{\langle \hat{m}\rangle}{m_c} + 5 \frac{\langle\hat{{\textsl k}}\hat{m}\rangle}{m_c}\right),
\end{split}
\end{equation}
where, ${\textsl k}_q \equiv \langle \hat{{\textsl k}}\rangle$. These expressions can be used to find the mean values of the macroscopic variables.

If we separate out the mean value from $\hat{{\textsl k}}$ as ${ \hat{{\textsl k}} = \hat{x} + \hat I} {\textsl k}_q$ then the error takes the simplified following form,

\begin{equation}
\frac{\sigma^{\rm Sys}_{\textsl k}}{\bar{{\textsl k}}} = \frac{\sqrt{\langle {\hat{x}^2} \rangle}}{\bar{{\textsl k}}},
\end{equation}
Note, a knowledge of the quantum state of the body will not only allow estimating ${\textsl k}_q$ but also $\sigma_{\textsl k}^{\rm Sys}$. Therefore if the systems do have quantum corrections, to measure its effect we have two observables to measure, namely the ${\textsl k}_q$ and $\sigma_{\textsl k}^{\rm Sys}$. Since in EMRIs statistical error will be low, it can help us infer the systematic error.

It is important to point out that Eq.(\ref{eq:delta-k}) is a model-dependent result found in Ref. \cite{Cardoso:2017cfl}. However, other models have found different scaling relations, such as Ref. \cite{Brustein:2020tpg} found $k\sim 1/|\ln\epsilon|^2$. Therefore, approaching the problem of probing quantum scales assuming a particular $\delta-k$ relation is not just. Rather, measuring $k_q$ and it's quantum systematic error can shed some light on the near-horizon quantum nature in a model-independent manner. It means that with EMRIs we can probe near horizon quantum scale larger than the Planck scale, making EMRIs the true GW microscopes.

Note, there is a degeneracy in the definition of ${\textsl k}$ \cite{Cardoso:2017cfl, Hinderer:2007mb, Binnington:2009bb}. Therefore depending on the definition of ${\textsl k}$, $\lambda \propto {\textsl k}_{CFMPR} m^5$ \cite{Cardoso:2017cfl} or $\lambda \propto {\textsl k}_{HBP} R^5$ \cite{Hinderer:2007mb, Binnington:2009bb}. In our work we considered the definition in Ref. \cite{Cardoso:2017cfl}, as connection with Planck scale physics is evident in this definition. However, most of the discussions in this work do not depend on one of the definitions. Therefore, while defining $\hat{{\textsl k}}$ this issue needs to be resolved. If the definition in Ref. \cite{Hinderer:2007mb, Binnington:2009bb} is considered then $\hat{m}$ will be replaced by $\hat{R}$ in the equations.

Using the prescription in this section we connect them with the observables. We have already argued that rather than focusing on any model dependent $\delta-k$ relation it is better to approach it in a model-agnostic manner. For that purpose, one should focus on measuring $k_q$ and $\sigma_{\textsl k}^{\rm Sys}$. During parameter estimation, the measurement of $k$ in this prescription will have both statistical and systematic error in a similar fashion as Eq. (\ref{systematic1}). Hence we can express it as follows:

\begin{equation}
\label{systematic}
\begin{split}
\frac{\sigma_{k}^{\rm Tot}}{\bar{k}} = \sqrt{\left(\frac{\sigma_{k}^{\rm Stat}}{\bar{k}}\right)^2 + \left(\frac{\sigma_{k}^{\rm Sys}}{\bar{k}}\right)^2}
\end{split}
\end{equation}
where, $\bar{k}$ is the estimated value of $k$ from the observation. ${\rm Stat}$ is the shorthand for statistical error just like before. 

As with EMRIs, the first term will become very small the error will be dominated by $\sigma_{k}^{\rm Sys}$ if $\sigma_{k}^{\rm Stat} \ll \sigma_{k}^{\rm Sys}$. In the context of $\delta-k$ relation, this was precisely the case as $a^2 \gtrsim 1$. Hence this can be used to infer $\sigma_{k}^{\rm Sys}$ or at least can be used to put some constraint on it.

This can be done by performing simulations with injected synthetic signal in detectors with $\bar{{\textsl k}}$. Running a Bayesian estimation on that we can have an estimation of $\sigma_{k}^{\rm Stat}$, which is an artifact of the observation. By estimating this value from simulations we can estimate the $\sigma_{k}^{\rm Sys}$, which is arising from the quantum nature. As there will be other sources of systematic error also, i.e. incomplete noise realization, and post-Newtonian truncation error to mention a few, we will only be able to put some upper bound on the quantum noise. This can lead us to understand the quantum states near the horizon.

\section{Discussion}
\label{Sec:discussion}

We have explored the resolving power of
the EMRIs as gravitational microscope which can be used to probe near horizon physics with TLN ${\textsl k}$. The presence of the environmental effects could impact the GW signal \cite{Yunes:2010sm, Yunes:2011ws, Kocsis:2011dr, Barausse:2014pra} and exclusion of them may lead to erroneous measurements of TLNs \cite{Cardoso:2019upw}. Similarly, other competing effects can also mimic the effect of tidal deformability \cite{Cardoso:2016rao, Cardoso:2016oxy, Datta:2019epe, Datta:2020rvo, Maggio:2021uge}. These should be taken into account to properly assess the potential of LISA. It is also required to study in detail from the theoretical standpoint the possible origin of these systems and their stability \cite{Addazi:2019bjz}.

We have explicitly shown for the first time that very small values of $k$ can add large dephasing in EMRIs. Our result suggests that it is possible for EMRIs to bring information regarding the quantum nature near the horizon scale. This paper also discusses the limitations of using the ECO relation between ${\textsl k}$ and $\delta$. We have also constructed a semi-classical formalism to take into account of the quantum effects. From the constructed formalism, it is evident that even if Eq. (\ref{eq:delta-k}) is not valid, there will be quantum signatures on the observables, at least in principle. We discussed how it should be estimated. To achieve our conclusions we have assumed the binary to be in an equatorial circular orbit, which is unlikely to be true for EMRIs. This should be investigated in the future.

Quantum effects for large astrophysical BHs are usually considered to be negligibly small. This conclusion arises from the expectation that the strength of quantum effects is governed by the ratio $\ell_{pl}^2 /r_s^2$. However, it was argued in Ref. \cite{Brustein:2020tpg} that the strength of quantum effects can be much larger because they can be governed by the ratio of $\ell_{pl}$ to the length scale of the fundamental theory of quantum gravity. In string theory, this is the string scale $l_s$. As a result, the quantum effects are governed by the ratio $g_s^2 = \ell_{pl}^2 /l_s^2$. Since $g_s^2$ can be $\sim 0.1$, it can have a larger contribution to the quantum effects \cite{Brustein:2020tpg}. This definitely requires further exploration.

Therefore it is high time to explore these avenues from the quantum gravity side. Finding possible effects of quantum gravity, as well as detailed numerical studies of coalescence of compact objects that has quantum contributions near their surfaces. This as a result will lead to proper quantification of quantum gravity effects on the GW observables.

{\bf Acknowledgement--} I would like to thank Bhaskar Biswas, Richard Brito, Sumanta Chakraborty, Paolo Pani, Niels Warburton, and Nicol$\acute{\rm a}$s Yunes for useful comments and also suggesting changes for the betterment of the article. I also thank Andrea Maselli, Swagat Mishra, Gabriel Andres Piovano, Karthik Rajeev, Shabbir Shaikh, and Yotam Sherf for useful discussions.
I would like to thank University Grants Commission (UGC), India, for financial support for a senior research fellowship.

\appendix

\section{Dephasing-Mismatch}
\label{mismatch}

To assess the strength of an effect to be measurable in a GW detector with  noise power spectral density $S_n(f)$, the overlap $\mathcal{O}$ between two waveforms $h_1(t)$ and $h_2(t)$ are usually computed:
\begin{equation}\label{overlap}
\mathcal{O}(h_1|h_2) = \frac{\left\langle h_1|h_2\right\rangle}{\sqrt{\left\langle h_1|h_1\right\rangle \left\langle h_2|h_2\right\rangle}}\,,
\end{equation}
where, the inner product $\left\langle h_1|h_2\right\rangle$ is defined as,
\begin{equation}
\left\langle h_1|h_2\right\rangle = 4\Re\,\int_{0}^{\infty} \frac{\tilde{h}_1 \tilde{h}^*_2}{S_n(f)} df\,.
\end{equation}
The quantities with tilde stand for the Fourier transform and the star for complex conjugation. As the waveforms 
are defined up to an arbitrary time and phase shift, it is required to maximize the overlap~\eqref{overlap} over 
these quantities. This can be done by computing~\cite{Allen:2005fk} 
\begin{equation}\label{overlap2}
\mathcal{O}(h_1|h_2) = \frac{4}{\sqrt{\left\langle h_1|h_1\right\rangle \left\langle h_2|h_2\right\rangle}}\max_{t_0} \left|\mathcal{F}^{-1}\left[\frac{\tilde{h}_1 \tilde{h}^*_2}{S_n(f)}\right](t_0)\right|\,,
\end{equation}
where $\mathcal{F}^{-1}[g(f)](t) =\int_{-\infty}^{+\infty} g(f) e^{-2\pi i f t}df$ is the inverse Fourier 
transform. The overlap is defined in such a manner that $\mathcal{O}=1$ indicates a perfect agreement between the two waveforms. The mismatch $(\mathfrak{M})$ is defined as follows:

\begin{equation}
    \mathfrak{M}\equiv 1-{\mathcal{O}}
\end{equation}

Two waveforms are considered to be indistinguishable for parameter estimation purposes if mismatch $\mathfrak{M} \lesssim 1/(2\rho^2)$~\cite{Flanagan:1997kp, Lindblom:2008cm}, where $\rho$ is the SNR of the true
signal. For an EMRI with an SNR $\rho \approx 20$ (resp., $\rho \approx 100)$ one has $\mathfrak{M}  \lesssim 10^{-3}$ (resp., $\mathfrak{M} \lesssim 5 \times 10^{-5})$. For a large number of parameters, say $D$, this relation gets slightly modified as $\mathfrak{M} \lesssim D/(2\rho^2)$\cite{Chatziioannou:2017tdw}.

Dephasing contribution $(\delta\phi)$ of an effect is indistinguishable from the absence of the effect in the context of scientific measurement if $\delta\phi^2 \lesssim 1/\rho^2 \sim \mathfrak{M}$. This condition is usually considered optimal in the sense that smaller dephasing than this is not measurable but not considering dephasing larger than this have distinguishable consequence \cite{Lindblom:2008cm}. The strongest LISA EMRIs may have SNR of up to $\rho \sim 100$ after matched filtering \cite{Amaro-Seoane:2007osp, Lindblom:2008cm, Porter:2010mb}, so phase differences on the order of $1/\rho$ radians should be just detectable in matched filtering \cite{Lindblom:2008cm, Amaro-Seoane:2011fgy, Katz:2021yft}. Keeping this in mind templates are constructed with $\delta\phi \leq 1/\rho.$ Therefore for SNR $\rho \approx 20 (100)$ $\mathfrak{M}  \sim 10^{-3}$ (resp., $\sim 5 \times 10^{-5})$ gets translated to dephasing $\delta \phi \approx .05(.01)$. This implies that for any reasonable SNR, dephasing $\delta\phi>1$ would eventually be detectable. In light of this, usually, it is conventional wisdom to consider $\delta\phi \sim 1$ radian as the detection threshold.

\bibliography{absorption.bib}

\begin{thebibliography}{91}%
\makeatletter
\providecommand \@ifxundefined [1]{%
 \@ifx{#1\undefined}
}%
\providecommand \@ifnum [1]{%
 \ifnum #1\expandafter \@firstoftwo
 \else \expandafter \@secondoftwo
 \fi
}%
\providecommand \@ifx [1]{%
 \ifx #1\expandafter \@firstoftwo
 \else \expandafter \@secondoftwo
 \fi
}%
\providecommand \natexlab [1]{#1}%
\providecommand \enquote  [1]{``#1''}%
\providecommand \bibnamefont  [1]{#1}%
\providecommand \bibfnamefont [1]{#1}%
\providecommand \citenamefont [1]{#1}%
\providecommand \href@noop [0]{\@secondoftwo}%
\providecommand \href [0]{\begingroup \@sanitize@url \@href}%
\providecommand \@href[1]{\@@startlink{#1}\@@href}%
\providecommand \@@href[1]{\endgroup#1\@@endlink}%
\providecommand \@sanitize@url [0]{\catcode `\\12\catcode `\$12\catcode
  `\&12\catcode `\#12\catcode `\^12\catcode `\_12\catcode `\%12\relax}%
\providecommand \@@startlink[1]{}%
\providecommand \@@endlink[0]{}%
\providecommand \url  [0]{\begingroup\@sanitize@url \@url }%
\providecommand \@url [1]{\endgroup\@href {#1}{\urlprefix }}%
\providecommand \urlprefix  [0]{URL }%
\providecommand \Eprint [0]{\href }%
\providecommand \doibase [0]{http://dx.doi.org/}%
\providecommand \selectlanguage [0]{\@gobble}%
\providecommand \bibinfo  [0]{\@secondoftwo}%
\providecommand \bibfield  [0]{\@secondoftwo}%
\providecommand \translation [1]{[#1]}%
\providecommand \BibitemOpen [0]{}%
\providecommand \bibitemStop [0]{}%
\providecommand \bibitemNoStop [0]{.\EOS\space}%
\providecommand \EOS [0]{\spacefactor3000\relax}%
\providecommand \BibitemShut  [1]{\csname bibitem#1\endcsname}%
\let\auto@bib@innerbib\@empty
\bibitem [{\citenamefont {Abbott}\ \emph
  {et~al.}(2016{\natexlab{a}})\citenamefont {Abbott} \emph
  {et~al.}}]{Abbott:2016blz}%
  \BibitemOpen
  \bibfield  {author} {\bibinfo {author} {\bibfnamefont {B.~P.}\ \bibnamefont
  {Abbott}} \emph {et~al.} (\bibinfo {collaboration} {LIGO Scientific,
  Virgo}),\ }\href {\doibase 10.1103/PhysRevLett.116.061102} {\bibfield
  {journal} {\bibinfo  {journal} {Phys. Rev. Lett.}\ }\textbf {\bibinfo
  {volume} {116}},\ \bibinfo {pages} {061102} (\bibinfo {year}
  {2016}{\natexlab{a}})},\ \Eprint {http://arxiv.org/abs/1602.03837}
  {arXiv:1602.03837 [gr-qc]} \BibitemShut {NoStop}%
\bibitem [{\citenamefont {Abbott}\ \emph
  {et~al.}(2019{\natexlab{a}})\citenamefont {Abbott} \emph
  {et~al.}}]{LIGOScientific:2018mvr}%
  \BibitemOpen
  \bibfield  {author} {\bibinfo {author} {\bibfnamefont {B.~P.}\ \bibnamefont
  {Abbott}} \emph {et~al.} (\bibinfo {collaboration} {LIGO Scientific,
  Virgo}),\ }\href {\doibase 10.1103/PhysRevX.9.031040} {\bibfield  {journal}
  {\bibinfo  {journal} {Phys. Rev.}\ }\textbf {\bibinfo {volume} {X9}},\
  \bibinfo {pages} {031040} (\bibinfo {year} {2019}{\natexlab{a}})},\ \Eprint
  {http://arxiv.org/abs/1811.12907} {arXiv:1811.12907 [astro-ph.HE]}
  \BibitemShut {NoStop}%
\bibitem [{\citenamefont {Abbott}\ \emph
  {et~al.}(2019{\natexlab{b}})\citenamefont {Abbott} \emph
  {et~al.}}]{LIGOScientific:2019fpa}%
  \BibitemOpen
  \bibfield  {author} {\bibinfo {author} {\bibfnamefont {B.~P.}\ \bibnamefont
  {Abbott}} \emph {et~al.} (\bibinfo {collaboration} {LIGO Scientific,
  Virgo}),\ }\href {\doibase 10.1103/PhysRevD.100.104036} {\bibfield  {journal}
  {\bibinfo  {journal} {Phys. Rev.}\ }\textbf {\bibinfo {volume} {D100}},\
  \bibinfo {pages} {104036} (\bibinfo {year} {2019}{\natexlab{b}})},\ \Eprint
  {http://arxiv.org/abs/1903.04467} {arXiv:1903.04467 [gr-qc]} \BibitemShut
  {NoStop}%
\bibitem [{\citenamefont {Abbott}\ \emph
  {et~al.}(2019{\natexlab{c}})\citenamefont {Abbott} \emph
  {et~al.}}]{Abbott:2018lct}%
  \BibitemOpen
  \bibfield  {author} {\bibinfo {author} {\bibfnamefont {B.~P.}\ \bibnamefont
  {Abbott}} \emph {et~al.} (\bibinfo {collaboration} {LIGO Scientific,
  Virgo}),\ }\href {\doibase 10.1103/PhysRevLett.123.011102} {\bibfield
  {journal} {\bibinfo  {journal} {Phys. Rev. Lett.}\ }\textbf {\bibinfo
  {volume} {123}},\ \bibinfo {pages} {011102} (\bibinfo {year}
  {2019}{\natexlab{c}})},\ \Eprint {http://arxiv.org/abs/1811.00364}
  {arXiv:1811.00364 [gr-qc]} \BibitemShut {NoStop}%
\bibitem [{\citenamefont {Abbott}\ \emph
  {et~al.}(2016{\natexlab{b}})\citenamefont {Abbott} \emph
  {et~al.}}]{TheLIGOScientific:2016pea}%
  \BibitemOpen
  \bibfield  {author} {\bibinfo {author} {\bibfnamefont {B.~P.}\ \bibnamefont
  {Abbott}} \emph {et~al.} (\bibinfo {collaboration} {LIGO Scientific,
  Virgo}),\ }\href {\doibase 10.1103/PhysRevX.6.041015,
  10.1103/PhysRevX.8.039903} {\bibfield  {journal} {\bibinfo  {journal} {Phys.
  Rev.}\ }\textbf {\bibinfo {volume} {X6}},\ \bibinfo {pages} {041015}
  (\bibinfo {year} {2016}{\natexlab{b}})},\ \bibinfo {note} {[erratum: Phys.
  Rev.X8,no.3,039903(2018)]},\ \Eprint {http://arxiv.org/abs/1606.04856}
  {arXiv:1606.04856 [gr-qc]} \BibitemShut {NoStop}%
\bibitem [{\citenamefont {Abbott}\ \emph
  {et~al.}(2016{\natexlab{c}})\citenamefont {Abbott} \emph
  {et~al.}}]{TheLIGOScientific:2016src}%
  \BibitemOpen
  \bibfield  {author} {\bibinfo {author} {\bibfnamefont {B.~P.}\ \bibnamefont
  {Abbott}} \emph {et~al.} (\bibinfo {collaboration} {LIGO Scientific,
  Virgo}),\ }\href {\doibase 10.1103/PhysRevLett.116.221101,
  10.1103/PhysRevLett.121.129902} {\bibfield  {journal} {\bibinfo  {journal}
  {Phys. Rev. Lett.}\ }\textbf {\bibinfo {volume} {116}},\ \bibinfo {pages}
  {221101} (\bibinfo {year} {2016}{\natexlab{c}})},\ \bibinfo {note} {[Erratum:
  Phys. Rev. Lett.121,no.12,129902(2018)]},\ \Eprint
  {http://arxiv.org/abs/1602.03841} {arXiv:1602.03841 [gr-qc]} \BibitemShut
  {NoStop}%
\bibitem [{\citenamefont {Abbott}\ \emph
  {et~al.}(2017{\natexlab{a}})\citenamefont {Abbott} \emph
  {et~al.}}]{Abbott:2017vtc}%
  \BibitemOpen
  \bibfield  {author} {\bibinfo {author} {\bibfnamefont {B.~P.}\ \bibnamefont
  {Abbott}} \emph {et~al.} (\bibinfo {collaboration} {LIGO Scientific,
  VIRGO}),\ }\href {\doibase 10.1103/PhysRevLett.118.221101,
  10.1103/PhysRevLett.121.129901} {\bibfield  {journal} {\bibinfo  {journal}
  {Phys. Rev. Lett.}\ }\textbf {\bibinfo {volume} {118}},\ \bibinfo {pages}
  {221101} (\bibinfo {year} {2017}{\natexlab{a}})},\ \bibinfo {note} {[Erratum:
  Phys. Rev. Lett.121,no.12,129901(2018)]},\ \Eprint
  {http://arxiv.org/abs/1706.01812} {arXiv:1706.01812 [gr-qc]} \BibitemShut
  {NoStop}%
\bibitem [{\citenamefont {Cardoso}\ \emph {et~al.}(2017)\citenamefont
  {Cardoso}, \citenamefont {Franzin}, \citenamefont {Maselli}, \citenamefont
  {Pani},\ and\ \citenamefont {Raposo}}]{Cardoso:2017cfl}%
  \BibitemOpen
  \bibfield  {author} {\bibinfo {author} {\bibfnamefont {V.}~\bibnamefont
  {Cardoso}}, \bibinfo {author} {\bibfnamefont {E.}~\bibnamefont {Franzin}},
  \bibinfo {author} {\bibfnamefont {A.}~\bibnamefont {Maselli}}, \bibinfo
  {author} {\bibfnamefont {P.}~\bibnamefont {Pani}}, \ and\ \bibinfo {author}
  {\bibfnamefont {G.}~\bibnamefont {Raposo}},\ }\href {\doibase
  10.1103/PhysRevD.95.084014} {\bibfield  {journal} {\bibinfo  {journal} {Phys.
  Rev. D}\ }\textbf {\bibinfo {volume} {95}},\ \bibinfo {pages} {084014}
  (\bibinfo {year} {2017})},\ \bibinfo {note} {[Addendum: Phys.Rev.D 95, 089901
  (2017)]},\ \Eprint {http://arxiv.org/abs/1701.01116} {arXiv:1701.01116
  [gr-qc]} \BibitemShut {NoStop}%
\bibitem [{\citenamefont {Sennett}\ \emph {et~al.}(2017)\citenamefont
  {Sennett}, \citenamefont {Hinderer}, \citenamefont {Steinhoff}, \citenamefont
  {Buonanno},\ and\ \citenamefont {Ossokine}}]{Sennett:2017etc}%
  \BibitemOpen
  \bibfield  {author} {\bibinfo {author} {\bibfnamefont {N.}~\bibnamefont
  {Sennett}}, \bibinfo {author} {\bibfnamefont {T.}~\bibnamefont {Hinderer}},
  \bibinfo {author} {\bibfnamefont {J.}~\bibnamefont {Steinhoff}}, \bibinfo
  {author} {\bibfnamefont {A.}~\bibnamefont {Buonanno}}, \ and\ \bibinfo
  {author} {\bibfnamefont {S.}~\bibnamefont {Ossokine}},\ }\href {\doibase
  10.1103/PhysRevD.96.024002} {\bibfield  {journal} {\bibinfo  {journal} {Phys.
  Rev.}\ }\textbf {\bibinfo {volume} {D96}},\ \bibinfo {pages} {024002}
  (\bibinfo {year} {2017})},\ \Eprint {http://arxiv.org/abs/1704.08651}
  {arXiv:1704.08651 [gr-qc]} \BibitemShut {NoStop}%
\bibitem [{\citenamefont {Maselli}\ \emph {et~al.}(2018)\citenamefont
  {Maselli}, \citenamefont {Pani}, \citenamefont {Cardoso}, \citenamefont
  {Abdelsalhin}, \citenamefont {Gualtieri},\ and\ \citenamefont
  {Ferrari}}]{Maselli:2017cmm}%
  \BibitemOpen
  \bibfield  {author} {\bibinfo {author} {\bibfnamefont {A.}~\bibnamefont
  {Maselli}}, \bibinfo {author} {\bibfnamefont {P.}~\bibnamefont {Pani}},
  \bibinfo {author} {\bibfnamefont {V.}~\bibnamefont {Cardoso}}, \bibinfo
  {author} {\bibfnamefont {T.}~\bibnamefont {Abdelsalhin}}, \bibinfo {author}
  {\bibfnamefont {L.}~\bibnamefont {Gualtieri}}, \ and\ \bibinfo {author}
  {\bibfnamefont {V.}~\bibnamefont {Ferrari}},\ }\href {\doibase
  10.1103/PhysRevLett.120.081101} {\bibfield  {journal} {\bibinfo  {journal}
  {Phys. Rev. Lett.}\ }\textbf {\bibinfo {volume} {120}},\ \bibinfo {pages}
  {081101} (\bibinfo {year} {2018})},\ \Eprint
  {http://arxiv.org/abs/1703.10612} {arXiv:1703.10612 [gr-qc]} \BibitemShut
  {NoStop}%
\bibitem [{\citenamefont {Brustein}\ and\ \citenamefont
  {Sherf}(2020)}]{Brustein:2020tpg}%
  \BibitemOpen
  \bibfield  {author} {\bibinfo {author} {\bibfnamefont {R.}~\bibnamefont
  {Brustein}}\ and\ \bibinfo {author} {\bibfnamefont {Y.}~\bibnamefont
  {Sherf}},\ }\href@noop {} {\  (\bibinfo {year} {2020})},\ \Eprint
  {http://arxiv.org/abs/2008.02738} {arXiv:2008.02738 [gr-qc]} \BibitemShut
  {NoStop}%
\bibitem [{\citenamefont {Datta}\ and\ \citenamefont
  {Bose}(2019)}]{Datta:2019euh}%
  \BibitemOpen
  \bibfield  {author} {\bibinfo {author} {\bibfnamefont {S.}~\bibnamefont
  {Datta}}\ and\ \bibinfo {author} {\bibfnamefont {S.}~\bibnamefont {Bose}},\
  }\href {\doibase 10.1103/PhysRevD.99.084001} {\bibfield  {journal} {\bibinfo
  {journal} {Phys. Rev.}\ }\textbf {\bibinfo {volume} {D99}},\ \bibinfo {pages}
  {084001} (\bibinfo {year} {2019})},\ \Eprint
  {http://arxiv.org/abs/1902.01723} {arXiv:1902.01723 [gr-qc]} \BibitemShut
  {NoStop}%
\bibitem [{\citenamefont {Datta}\ \emph
  {et~al.}(2020{\natexlab{a}})\citenamefont {Datta}, \citenamefont {Brito},
  \citenamefont {Bose}, \citenamefont {Pani},\ and\ \citenamefont
  {Hughes}}]{Datta:2019epe}%
  \BibitemOpen
  \bibfield  {author} {\bibinfo {author} {\bibfnamefont {S.}~\bibnamefont
  {Datta}}, \bibinfo {author} {\bibfnamefont {R.}~\bibnamefont {Brito}},
  \bibinfo {author} {\bibfnamefont {S.}~\bibnamefont {Bose}}, \bibinfo {author}
  {\bibfnamefont {P.}~\bibnamefont {Pani}}, \ and\ \bibinfo {author}
  {\bibfnamefont {S.~A.}\ \bibnamefont {Hughes}},\ }\href {\doibase
  10.1103/PhysRevD.101.044004} {\bibfield  {journal} {\bibinfo  {journal}
  {Phys. Rev.}\ }\textbf {\bibinfo {volume} {D101}},\ \bibinfo {pages} {044004}
  (\bibinfo {year} {2020}{\natexlab{a}})},\ \Eprint
  {http://arxiv.org/abs/1910.07841} {arXiv:1910.07841 [gr-qc]} \BibitemShut
  {NoStop}%
\bibitem [{\citenamefont {Datta}(2020)}]{Datta:2020rvo}%
  \BibitemOpen
  \bibfield  {author} {\bibinfo {author} {\bibfnamefont {S.}~\bibnamefont
  {Datta}},\ }\href {\doibase 10.1103/PhysRevD.102.064040} {\bibfield
  {journal} {\bibinfo  {journal} {Phys. Rev. D}\ }\textbf {\bibinfo {volume}
  {102}},\ \bibinfo {pages} {064040} (\bibinfo {year} {2020})},\ \Eprint
  {http://arxiv.org/abs/2002.04480} {arXiv:2002.04480 [gr-qc]} \BibitemShut
  {NoStop}%
\bibitem [{\citenamefont {Agullo}\ \emph {et~al.}(2021)\citenamefont {Agullo},
  \citenamefont {Cardoso}, \citenamefont {Rio}, \citenamefont {Maggiore},\ and\
  \citenamefont {Pullin}}]{Agullo:2020hxe}%
  \BibitemOpen
  \bibfield  {author} {\bibinfo {author} {\bibfnamefont {I.}~\bibnamefont
  {Agullo}}, \bibinfo {author} {\bibfnamefont {V.}~\bibnamefont {Cardoso}},
  \bibinfo {author} {\bibfnamefont {A.~D.}\ \bibnamefont {Rio}}, \bibinfo
  {author} {\bibfnamefont {M.}~\bibnamefont {Maggiore}}, \ and\ \bibinfo
  {author} {\bibfnamefont {J.}~\bibnamefont {Pullin}},\ }\href {\doibase
  10.1103/PhysRevLett.126.041302} {\bibfield  {journal} {\bibinfo  {journal}
  {Phys. Rev. Lett.}\ }\textbf {\bibinfo {volume} {126}},\ \bibinfo {pages}
  {041302} (\bibinfo {year} {2021})},\ \Eprint
  {http://arxiv.org/abs/2007.13761} {arXiv:2007.13761 [gr-qc]} \BibitemShut
  {NoStop}%
\bibitem [{\citenamefont {Chakraborty}\ \emph {et~al.}(2021)\citenamefont
  {Chakraborty}, \citenamefont {Datta},\ and\ \citenamefont
  {Sau}}]{Chakraborty:2021gdf}%
  \BibitemOpen
  \bibfield  {author} {\bibinfo {author} {\bibfnamefont {S.}~\bibnamefont
  {Chakraborty}}, \bibinfo {author} {\bibfnamefont {S.}~\bibnamefont {Datta}},
  \ and\ \bibinfo {author} {\bibfnamefont {S.}~\bibnamefont {Sau}},\
  }\href@noop {} {\  (\bibinfo {year} {2021})},\ \Eprint
  {http://arxiv.org/abs/2103.12430} {arXiv:2103.12430 [gr-qc]} \BibitemShut
  {NoStop}%
\bibitem [{\citenamefont {Sherf}(2021)}]{Sherf:2021ppp}%
  \BibitemOpen
  \bibfield  {author} {\bibinfo {author} {\bibfnamefont {Y.}~\bibnamefont
  {Sherf}},\ }\href {\doibase 10.1103/PhysRevD.103.104003} {\bibfield
  {journal} {\bibinfo  {journal} {Phys. Rev. D}\ }\textbf {\bibinfo {volume}
  {103}},\ \bibinfo {pages} {104003} (\bibinfo {year} {2021})},\ \Eprint
  {http://arxiv.org/abs/2104.03766} {arXiv:2104.03766 [gr-qc]} \BibitemShut
  {NoStop}%
\bibitem [{\citenamefont {Datta}\ and\ \citenamefont
  {Phukon}(2021)}]{Datta:2021row}%
  \BibitemOpen
  \bibfield  {author} {\bibinfo {author} {\bibfnamefont {S.}~\bibnamefont
  {Datta}}\ and\ \bibinfo {author} {\bibfnamefont {K.~S.}\ \bibnamefont
  {Phukon}},\ }\href@noop {} {\  (\bibinfo {year} {2021})},\ \Eprint
  {http://arxiv.org/abs/2105.11140} {arXiv:2105.11140 [gr-qc]} \BibitemShut
  {NoStop}%
\bibitem [{\citenamefont {Sago}\ and\ \citenamefont
  {Tanaka}(2021)}]{Sago:2021iku}%
  \BibitemOpen
  \bibfield  {author} {\bibinfo {author} {\bibfnamefont {N.}~\bibnamefont
  {Sago}}\ and\ \bibinfo {author} {\bibfnamefont {T.}~\bibnamefont {Tanaka}},\
  }\href {\doibase 10.1103/PhysRevD.104.064009} {\bibfield  {journal} {\bibinfo
   {journal} {Phys. Rev. D}\ }\textbf {\bibinfo {volume} {104}},\ \bibinfo
  {pages} {064009} (\bibinfo {year} {2021})},\ \Eprint
  {http://arxiv.org/abs/2106.07123} {arXiv:2106.07123 [gr-qc]} \BibitemShut
  {NoStop}%
\bibitem [{\citenamefont {Maggio}\ \emph {et~al.}(2021)\citenamefont {Maggio},
  \citenamefont {van~de Meent},\ and\ \citenamefont {Pani}}]{Maggio:2021uge}%
  \BibitemOpen
  \bibfield  {author} {\bibinfo {author} {\bibfnamefont {E.}~\bibnamefont
  {Maggio}}, \bibinfo {author} {\bibfnamefont {M.}~\bibnamefont {van~de
  Meent}}, \ and\ \bibinfo {author} {\bibfnamefont {P.}~\bibnamefont {Pani}},\
  }\href@noop {} {\  (\bibinfo {year} {2021})},\ \Eprint
  {http://arxiv.org/abs/2106.07195} {arXiv:2106.07195 [gr-qc]} \BibitemShut
  {NoStop}%
\bibitem [{\citenamefont {Sago}\ and\ \citenamefont
  {Tanaka}(2022)}]{Sago:2022bbj}%
  \BibitemOpen
  \bibfield  {author} {\bibinfo {author} {\bibfnamefont {N.}~\bibnamefont
  {Sago}}\ and\ \bibinfo {author} {\bibfnamefont {T.}~\bibnamefont {Tanaka}},\
  }\href@noop {} {\  (\bibinfo {year} {2022})},\ \Eprint
  {http://arxiv.org/abs/2202.04249} {arXiv:2202.04249 [gr-qc]} \BibitemShut
  {NoStop}%
\bibitem [{\citenamefont {Mukherjee}\ \emph {et~al.}(2022)\citenamefont
  {Mukherjee}, \citenamefont {Datta}, \citenamefont {Tiwari}, \citenamefont
  {Phukon},\ and\ \citenamefont {Bose}}]{Mukherjee:2022wws}%
  \BibitemOpen
  \bibfield  {author} {\bibinfo {author} {\bibfnamefont {S.}~\bibnamefont
  {Mukherjee}}, \bibinfo {author} {\bibfnamefont {S.}~\bibnamefont {Datta}},
  \bibinfo {author} {\bibfnamefont {S.}~\bibnamefont {Tiwari}}, \bibinfo
  {author} {\bibfnamefont {K.~S.}\ \bibnamefont {Phukon}}, \ and\ \bibinfo
  {author} {\bibfnamefont {S.}~\bibnamefont {Bose}},\ }\href@noop {} {\
  (\bibinfo {year} {2022})},\ \Eprint {http://arxiv.org/abs/2202.08661}
  {arXiv:2202.08661 [gr-qc]} \BibitemShut {NoStop}%
\bibitem [{\citenamefont {Krishnendu}\ \emph {et~al.}(2017)\citenamefont
  {Krishnendu}, \citenamefont {Arun},\ and\ \citenamefont
  {Mishra}}]{Krishnendu:2017shb}%
  \BibitemOpen
  \bibfield  {author} {\bibinfo {author} {\bibfnamefont {N.~V.}\ \bibnamefont
  {Krishnendu}}, \bibinfo {author} {\bibfnamefont {K.~G.}\ \bibnamefont
  {Arun}}, \ and\ \bibinfo {author} {\bibfnamefont {C.~K.}\ \bibnamefont
  {Mishra}},\ }\href {\doibase 10.1103/PhysRevLett.119.091101} {\bibfield
  {journal} {\bibinfo  {journal} {Phys. Rev. Lett.}\ }\textbf {\bibinfo
  {volume} {119}},\ \bibinfo {pages} {091101} (\bibinfo {year} {2017})},\
  \Eprint {http://arxiv.org/abs/1701.06318} {arXiv:1701.06318 [gr-qc]}
  \BibitemShut {NoStop}%
\bibitem [{\citenamefont {Bianchi}\ \emph {et~al.}(2020)\citenamefont
  {Bianchi}, \citenamefont {Consoli}, \citenamefont {Grillo}, \citenamefont
  {Morales}, \citenamefont {Pani},\ and\ \citenamefont
  {Raposo}}]{Bianchi:2020bxa}%
  \BibitemOpen
  \bibfield  {author} {\bibinfo {author} {\bibfnamefont {M.}~\bibnamefont
  {Bianchi}}, \bibinfo {author} {\bibfnamefont {D.}~\bibnamefont {Consoli}},
  \bibinfo {author} {\bibfnamefont {A.}~\bibnamefont {Grillo}}, \bibinfo
  {author} {\bibfnamefont {J.~F.}\ \bibnamefont {Morales}}, \bibinfo {author}
  {\bibfnamefont {P.}~\bibnamefont {Pani}}, \ and\ \bibinfo {author}
  {\bibfnamefont {G.}~\bibnamefont {Raposo}},\ }\href {\doibase
  10.1103/PhysRevLett.125.221601} {\bibfield  {journal} {\bibinfo  {journal}
  {Phys. Rev. Lett.}\ }\textbf {\bibinfo {volume} {125}},\ \bibinfo {pages}
  {221601} (\bibinfo {year} {2020})},\ \Eprint
  {http://arxiv.org/abs/2007.01743} {arXiv:2007.01743 [hep-th]} \BibitemShut
  {NoStop}%
\bibitem [{\citenamefont {Mukherjee}\ and\ \citenamefont
  {Chakraborty}(2020)}]{Mukherjee:2020how}%
  \BibitemOpen
  \bibfield  {author} {\bibinfo {author} {\bibfnamefont {S.}~\bibnamefont
  {Mukherjee}}\ and\ \bibinfo {author} {\bibfnamefont {S.}~\bibnamefont
  {Chakraborty}},\ }\href {\doibase 10.1103/PhysRevD.102.124058} {\bibfield
  {journal} {\bibinfo  {journal} {Phys. Rev. D}\ }\textbf {\bibinfo {volume}
  {102}},\ \bibinfo {pages} {124058} (\bibinfo {year} {2020})},\ \Eprint
  {http://arxiv.org/abs/2008.06891} {arXiv:2008.06891 [gr-qc]} \BibitemShut
  {NoStop}%
\bibitem [{\citenamefont {Datta}\ and\ \citenamefont
  {Mukherjee}(2021)}]{Datta:2020axm}%
  \BibitemOpen
  \bibfield  {author} {\bibinfo {author} {\bibfnamefont {S.}~\bibnamefont
  {Datta}}\ and\ \bibinfo {author} {\bibfnamefont {S.}~\bibnamefont
  {Mukherjee}},\ }\href {\doibase 10.1103/PhysRevD.103.104032} {\bibfield
  {journal} {\bibinfo  {journal} {Phys. Rev. D}\ }\textbf {\bibinfo {volume}
  {103}},\ \bibinfo {pages} {104032} (\bibinfo {year} {2021})},\ \Eprint
  {http://arxiv.org/abs/2010.12387} {arXiv:2010.12387 [gr-qc]} \BibitemShut
  {NoStop}%
\bibitem [{\citenamefont {Narikawa}\ \emph {et~al.}(2021)\citenamefont
  {Narikawa}, \citenamefont {Uchikata},\ and\ \citenamefont
  {Tanaka}}]{Narikawa:2021pak}%
  \BibitemOpen
  \bibfield  {author} {\bibinfo {author} {\bibfnamefont {T.}~\bibnamefont
  {Narikawa}}, \bibinfo {author} {\bibfnamefont {N.}~\bibnamefont {Uchikata}},
  \ and\ \bibinfo {author} {\bibfnamefont {T.}~\bibnamefont {Tanaka}},\
  }\href@noop {} {\  (\bibinfo {year} {2021})},\ \Eprint
  {http://arxiv.org/abs/2106.09193} {arXiv:2106.09193 [gr-qc]} \BibitemShut
  {NoStop}%
\bibitem [{\citenamefont {Cardoso}\ \emph
  {et~al.}(2016{\natexlab{a}})\citenamefont {Cardoso}, \citenamefont
  {Franzin},\ and\ \citenamefont {Pani}}]{Cardoso:2016rao}%
  \BibitemOpen
  \bibfield  {author} {\bibinfo {author} {\bibfnamefont {V.}~\bibnamefont
  {Cardoso}}, \bibinfo {author} {\bibfnamefont {E.}~\bibnamefont {Franzin}}, \
  and\ \bibinfo {author} {\bibfnamefont {P.}~\bibnamefont {Pani}},\ }\href
  {\doibase 10.1103/PhysRevLett.116.171101} {\bibfield  {journal} {\bibinfo
  {journal} {Phys. Rev. Lett.}\ }\textbf {\bibinfo {volume} {116}},\ \bibinfo
  {pages} {171101} (\bibinfo {year} {2016}{\natexlab{a}})},\ \bibinfo {note}
  {[Erratum: Phys.Rev.Lett. 117, 089902 (2016)]},\ \Eprint
  {http://arxiv.org/abs/1602.07309} {arXiv:1602.07309 [gr-qc]} \BibitemShut
  {NoStop}%
\bibitem [{\citenamefont {Cardoso}\ \emph
  {et~al.}(2016{\natexlab{b}})\citenamefont {Cardoso}, \citenamefont {Hopper},
  \citenamefont {Macedo}, \citenamefont {Palenzuela},\ and\ \citenamefont
  {Pani}}]{Cardoso:2016oxy}%
  \BibitemOpen
  \bibfield  {author} {\bibinfo {author} {\bibfnamefont {V.}~\bibnamefont
  {Cardoso}}, \bibinfo {author} {\bibfnamefont {S.}~\bibnamefont {Hopper}},
  \bibinfo {author} {\bibfnamefont {C.~F.~B.}\ \bibnamefont {Macedo}}, \bibinfo
  {author} {\bibfnamefont {C.}~\bibnamefont {Palenzuela}}, \ and\ \bibinfo
  {author} {\bibfnamefont {P.}~\bibnamefont {Pani}},\ }\href {\doibase
  10.1103/PhysRevD.94.084031} {\bibfield  {journal} {\bibinfo  {journal} {Phys.
  Rev.}\ }\textbf {\bibinfo {volume} {D94}},\ \bibinfo {pages} {084031}
  (\bibinfo {year} {2016}{\natexlab{b}})},\ \Eprint
  {http://arxiv.org/abs/1608.08637} {arXiv:1608.08637 [gr-qc]} \BibitemShut
  {NoStop}%
\bibitem [{\citenamefont {Maggio}\ \emph {et~al.}(2019)\citenamefont {Maggio},
  \citenamefont {Testa}, \citenamefont {Bhagwat},\ and\ \citenamefont
  {Pani}}]{Maggio:2019zyv}%
  \BibitemOpen
  \bibfield  {author} {\bibinfo {author} {\bibfnamefont {E.}~\bibnamefont
  {Maggio}}, \bibinfo {author} {\bibfnamefont {A.}~\bibnamefont {Testa}},
  \bibinfo {author} {\bibfnamefont {S.}~\bibnamefont {Bhagwat}}, \ and\
  \bibinfo {author} {\bibfnamefont {P.}~\bibnamefont {Pani}},\ }\href {\doibase
  10.1103/PhysRevD.100.064056} {\bibfield  {journal} {\bibinfo  {journal}
  {Phys. Rev. D}\ }\textbf {\bibinfo {volume} {100}},\ \bibinfo {pages}
  {064056} (\bibinfo {year} {2019})},\ \Eprint
  {http://arxiv.org/abs/1907.03091} {arXiv:1907.03091 [gr-qc]} \BibitemShut
  {NoStop}%
\bibitem [{\citenamefont {Tsang}\ \emph {et~al.}(2020)\citenamefont {Tsang},
  \citenamefont {Ghosh}, \citenamefont {Samajdar}, \citenamefont
  {Chatziioannou}, \citenamefont {Mastrogiovanni}, \citenamefont {Agathos},\
  and\ \citenamefont {Van Den~Broeck}}]{Tsang:2019zra}%
  \BibitemOpen
  \bibfield  {author} {\bibinfo {author} {\bibfnamefont {K.~W.}\ \bibnamefont
  {Tsang}}, \bibinfo {author} {\bibfnamefont {A.}~\bibnamefont {Ghosh}},
  \bibinfo {author} {\bibfnamefont {A.}~\bibnamefont {Samajdar}}, \bibinfo
  {author} {\bibfnamefont {K.}~\bibnamefont {Chatziioannou}}, \bibinfo {author}
  {\bibfnamefont {S.}~\bibnamefont {Mastrogiovanni}}, \bibinfo {author}
  {\bibfnamefont {M.}~\bibnamefont {Agathos}}, \ and\ \bibinfo {author}
  {\bibfnamefont {C.}~\bibnamefont {Van Den~Broeck}},\ }\href {\doibase
  10.1103/PhysRevD.101.064012} {\bibfield  {journal} {\bibinfo  {journal}
  {Phys.\ Rev.\ D}\ }\textbf {\bibinfo {volume} {101}},\ \bibinfo {pages}
  {064012} (\bibinfo {year} {2020})},\ \Eprint
  {http://arxiv.org/abs/1906.11168} {arXiv:1906.11168 [gr-qc]} \BibitemShut
  {NoStop}%
\bibitem [{\citenamefont {Abedi}\ \emph {et~al.}(2017)\citenamefont {Abedi},
  \citenamefont {Dykaar},\ and\ \citenamefont {Afshordi}}]{Abedi:2016hgu}%
  \BibitemOpen
  \bibfield  {author} {\bibinfo {author} {\bibfnamefont {J.}~\bibnamefont
  {Abedi}}, \bibinfo {author} {\bibfnamefont {H.}~\bibnamefont {Dykaar}}, \
  and\ \bibinfo {author} {\bibfnamefont {N.}~\bibnamefont {Afshordi}},\ }\href
  {\doibase 10.1103/PhysRevD.96.082004} {\bibfield  {journal} {\bibinfo
  {journal} {Phys. Rev.}\ }\textbf {\bibinfo {volume} {D96}},\ \bibinfo {pages}
  {082004} (\bibinfo {year} {2017})},\ \Eprint
  {http://arxiv.org/abs/1612.00266} {arXiv:1612.00266 [gr-qc]} \BibitemShut
  {NoStop}%
\bibitem [{\citenamefont {Westerweck}\ \emph {et~al.}(2018)\citenamefont
  {Westerweck}, \citenamefont {Nielsen}, \citenamefont {Fischer-Birnholtz},
  \citenamefont {Cabero}, \citenamefont {Capano}, \citenamefont {Dent},
  \citenamefont {Krishnan}, \citenamefont {Meadors},\ and\ \citenamefont
  {Nitz}}]{Westerweck:2017hus}%
  \BibitemOpen
  \bibfield  {author} {\bibinfo {author} {\bibfnamefont {J.}~\bibnamefont
  {Westerweck}}, \bibinfo {author} {\bibfnamefont {A.}~\bibnamefont {Nielsen}},
  \bibinfo {author} {\bibfnamefont {O.}~\bibnamefont {Fischer-Birnholtz}},
  \bibinfo {author} {\bibfnamefont {M.}~\bibnamefont {Cabero}}, \bibinfo
  {author} {\bibfnamefont {C.}~\bibnamefont {Capano}}, \bibinfo {author}
  {\bibfnamefont {T.}~\bibnamefont {Dent}}, \bibinfo {author} {\bibfnamefont
  {B.}~\bibnamefont {Krishnan}}, \bibinfo {author} {\bibfnamefont
  {G.}~\bibnamefont {Meadors}}, \ and\ \bibinfo {author} {\bibfnamefont
  {A.~H.}\ \bibnamefont {Nitz}},\ }\href {\doibase 10.1103/PhysRevD.97.124037}
  {\bibfield  {journal} {\bibinfo  {journal} {Phys. Rev.}\ }\textbf {\bibinfo
  {volume} {D97}},\ \bibinfo {pages} {124037} (\bibinfo {year} {2018})},\
  \Eprint {http://arxiv.org/abs/1712.09966} {arXiv:1712.09966 [gr-qc]}
  \BibitemShut {NoStop}%
\bibitem [{\citenamefont {Cardoso}\ and\ \citenamefont
  {Pani}(2019)}]{Cardoso:2019rvt}%
  \BibitemOpen
  \bibfield  {author} {\bibinfo {author} {\bibfnamefont {V.}~\bibnamefont
  {Cardoso}}\ and\ \bibinfo {author} {\bibfnamefont {P.}~\bibnamefont {Pani}},\
  }\href@noop {} {\  (\bibinfo {year} {2019})},\ \Eprint
  {http://arxiv.org/abs/1904.05363} {arXiv:1904.05363 [gr-qc]} \BibitemShut
  {NoStop}%
\bibitem [{\citenamefont {Chen}\ \emph {et~al.}(2020)\citenamefont {Chen},
  \citenamefont {Wang},\ and\ \citenamefont {Chen}}]{Chen:2020htz}%
  \BibitemOpen
  \bibfield  {author} {\bibinfo {author} {\bibfnamefont {B.}~\bibnamefont
  {Chen}}, \bibinfo {author} {\bibfnamefont {Q.}~\bibnamefont {Wang}}, \ and\
  \bibinfo {author} {\bibfnamefont {Y.}~\bibnamefont {Chen}},\ }\href@noop {}
  {\  (\bibinfo {year} {2020})},\ \Eprint {http://arxiv.org/abs/2012.10842}
  {arXiv:2012.10842 [gr-qc]} \BibitemShut {NoStop}%
\bibitem [{\citenamefont {Xin}\ \emph {et~al.}(2021)\citenamefont {Xin},
  \citenamefont {Chen}, \citenamefont {Lo}, \citenamefont {Sun}, \citenamefont
  {Han}, \citenamefont {Zhong}, \citenamefont {Srivastava}, \citenamefont {Ma},
  \citenamefont {Wang},\ and\ \citenamefont {Chen}}]{Xin:2021zir}%
  \BibitemOpen
  \bibfield  {author} {\bibinfo {author} {\bibfnamefont {S.}~\bibnamefont
  {Xin}}, \bibinfo {author} {\bibfnamefont {B.}~\bibnamefont {Chen}}, \bibinfo
  {author} {\bibfnamefont {R.~K.~L.}\ \bibnamefont {Lo}}, \bibinfo {author}
  {\bibfnamefont {L.}~\bibnamefont {Sun}}, \bibinfo {author} {\bibfnamefont
  {W.-B.}\ \bibnamefont {Han}}, \bibinfo {author} {\bibfnamefont
  {X.}~\bibnamefont {Zhong}}, \bibinfo {author} {\bibfnamefont
  {M.}~\bibnamefont {Srivastava}}, \bibinfo {author} {\bibfnamefont
  {S.}~\bibnamefont {Ma}}, \bibinfo {author} {\bibfnamefont {Q.}~\bibnamefont
  {Wang}}, \ and\ \bibinfo {author} {\bibfnamefont {Y.}~\bibnamefont {Chen}},\
  }\href@noop {} {\  (\bibinfo {year} {2021})},\ \Eprint
  {http://arxiv.org/abs/2105.12313} {arXiv:2105.12313 [gr-qc]} \BibitemShut
  {NoStop}%
\bibitem [{\citenamefont {Titarchuk}\ and\ \citenamefont
  {Shaposhnikov}(2005)}]{Titarchuk:2005rr}%
  \BibitemOpen
  \bibfield  {author} {\bibinfo {author} {\bibfnamefont {L.}~\bibnamefont
  {Titarchuk}}\ and\ \bibinfo {author} {\bibfnamefont {N.}~\bibnamefont
  {Shaposhnikov}},\ }\href {\doibase 10.1086/429986} {\bibfield  {journal}
  {\bibinfo  {journal} {Astrophys. J.}\ }\textbf {\bibinfo {volume} {626}},\
  \bibinfo {pages} {298} (\bibinfo {year} {2005})},\ \Eprint
  {http://arxiv.org/abs/astro-ph/0503081} {arXiv:astro-ph/0503081} \BibitemShut
  {NoStop}%
\bibitem [{\citenamefont {Bambi}(2013)}]{Bambi:2013sha}%
  \BibitemOpen
  \bibfield  {author} {\bibinfo {author} {\bibfnamefont {C.}~\bibnamefont
  {Bambi}},\ }\href {\doibase 10.1088/1475-7516/2013/08/055} {\bibfield
  {journal} {\bibinfo  {journal} {JCAP}\ }\textbf {\bibinfo {volume} {08}},\
  \bibinfo {pages} {055} (\bibinfo {year} {2013})},\ \Eprint
  {http://arxiv.org/abs/1305.5409} {arXiv:1305.5409 [gr-qc]} \BibitemShut
  {NoStop}%
\bibitem [{\citenamefont {Jiang}\ \emph {et~al.}(2015)\citenamefont {Jiang},
  \citenamefont {Bambi},\ and\ \citenamefont {Steiner}}]{Jiang:2014loa}%
  \BibitemOpen
  \bibfield  {author} {\bibinfo {author} {\bibfnamefont {J.}~\bibnamefont
  {Jiang}}, \bibinfo {author} {\bibfnamefont {C.}~\bibnamefont {Bambi}}, \ and\
  \bibinfo {author} {\bibfnamefont {J.~F.}\ \bibnamefont {Steiner}},\ }\href
  {\doibase 10.1088/1475-7516/2015/05/025} {\bibfield  {journal} {\bibinfo
  {journal} {JCAP}\ }\textbf {\bibinfo {volume} {05}},\ \bibinfo {pages} {025}
  (\bibinfo {year} {2015})},\ \Eprint {http://arxiv.org/abs/1406.5677}
  {arXiv:1406.5677 [gr-qc]} \BibitemShut {NoStop}%
\bibitem [{\citenamefont {Bambi}(2017)}]{Bambi:2015kza}%
  \BibitemOpen
  \bibfield  {author} {\bibinfo {author} {\bibfnamefont {C.}~\bibnamefont
  {Bambi}},\ }\href {\doibase 10.1103/RevModPhys.89.025001} {\bibfield
  {journal} {\bibinfo  {journal} {Rev. Mod. Phys.}\ }\textbf {\bibinfo {volume}
  {89}},\ \bibinfo {pages} {025001} (\bibinfo {year} {2017})},\ \Eprint
  {http://arxiv.org/abs/1509.03884} {arXiv:1509.03884 [gr-qc]} \BibitemShut
  {NoStop}%
\bibitem [{\citenamefont {Jenkins}\ \emph {et~al.}(2018)\citenamefont
  {Jenkins}, \citenamefont {Pithis},\ and\ \citenamefont
  {Sakellariadou}}]{Jenkins:2018ysa}%
  \BibitemOpen
  \bibfield  {author} {\bibinfo {author} {\bibfnamefont {A.~C.}\ \bibnamefont
  {Jenkins}}, \bibinfo {author} {\bibfnamefont {A.~G.~A.}\ \bibnamefont
  {Pithis}}, \ and\ \bibinfo {author} {\bibfnamefont {M.}~\bibnamefont
  {Sakellariadou}},\ }\href {\doibase 10.1103/PhysRevD.98.104032} {\bibfield
  {journal} {\bibinfo  {journal} {Phys. Rev. D}\ }\textbf {\bibinfo {volume}
  {98}},\ \bibinfo {pages} {104032} (\bibinfo {year} {2018})},\ \Eprint
  {http://arxiv.org/abs/1809.06275} {arXiv:1809.06275 [gr-qc]} \BibitemShut
  {NoStop}%
\bibitem [{\citenamefont {Calmet}\ \emph {et~al.}(2018)\citenamefont {Calmet},
  \citenamefont {El-Menoufi}, \citenamefont {Latosh},\ and\ \citenamefont
  {Mohapatra}}]{Calmet:2018rkj}%
  \BibitemOpen
  \bibfield  {author} {\bibinfo {author} {\bibfnamefont {X.}~\bibnamefont
  {Calmet}}, \bibinfo {author} {\bibfnamefont {B.~K.}\ \bibnamefont
  {El-Menoufi}}, \bibinfo {author} {\bibfnamefont {B.}~\bibnamefont {Latosh}},
  \ and\ \bibinfo {author} {\bibfnamefont {S.}~\bibnamefont {Mohapatra}},\
  }\href {\doibase 10.1140/epjc/s10052-018-6265-3} {\bibfield  {journal}
  {\bibinfo  {journal} {Eur. Phys. J. C}\ }\textbf {\bibinfo {volume} {78}},\
  \bibinfo {pages} {780} (\bibinfo {year} {2018})},\ \Eprint
  {http://arxiv.org/abs/1809.07606} {arXiv:1809.07606 [hep-th]} \BibitemShut
  {NoStop}%
\bibitem [{\citenamefont {Cardoso}\ \emph {et~al.}(2019)\citenamefont
  {Cardoso}, \citenamefont {Foit},\ and\ \citenamefont
  {Kleban}}]{Cardoso:2019apo}%
  \BibitemOpen
  \bibfield  {author} {\bibinfo {author} {\bibfnamefont {V.}~\bibnamefont
  {Cardoso}}, \bibinfo {author} {\bibfnamefont {V.~F.}\ \bibnamefont {Foit}}, \
  and\ \bibinfo {author} {\bibfnamefont {M.}~\bibnamefont {Kleban}},\ }\href
  {\doibase 10.1088/1475-7516/2019/08/006} {\bibfield  {journal} {\bibinfo
  {journal} {JCAP}\ }\textbf {\bibinfo {volume} {08}},\ \bibinfo {pages} {006}
  (\bibinfo {year} {2019})},\ \Eprint {http://arxiv.org/abs/1902.10164}
  {arXiv:1902.10164 [hep-th]} \BibitemShut {NoStop}%
\bibitem [{\citenamefont {Laghi}\ \emph {et~al.}(2021)\citenamefont {Laghi},
  \citenamefont {Carullo}, \citenamefont {Veitch},\ and\ \citenamefont
  {Del~Pozzo}}]{Laghi:2020rgl}%
  \BibitemOpen
  \bibfield  {author} {\bibinfo {author} {\bibfnamefont {D.}~\bibnamefont
  {Laghi}}, \bibinfo {author} {\bibfnamefont {G.}~\bibnamefont {Carullo}},
  \bibinfo {author} {\bibfnamefont {J.}~\bibnamefont {Veitch}}, \ and\ \bibinfo
  {author} {\bibfnamefont {W.}~\bibnamefont {Del~Pozzo}},\ }\href {\doibase
  10.1088/1361-6382/abde19} {\bibfield  {journal} {\bibinfo  {journal} {Class.
  Quant. Grav.}\ }\textbf {\bibinfo {volume} {38}},\ \bibinfo {pages} {095005}
  (\bibinfo {year} {2021})},\ \Eprint {http://arxiv.org/abs/2011.03816}
  {arXiv:2011.03816 [gr-qc]} \BibitemShut {NoStop}%
\bibitem [{\citenamefont {Brustein}\ and\ \citenamefont
  {Sherf}(2021)}]{Brustein:2021bnw}%
  \BibitemOpen
  \bibfield  {author} {\bibinfo {author} {\bibfnamefont {R.}~\bibnamefont
  {Brustein}}\ and\ \bibinfo {author} {\bibfnamefont {Y.}~\bibnamefont
  {Sherf}},\ }\href@noop {} {\  (\bibinfo {year} {2021})},\ \Eprint
  {http://arxiv.org/abs/2104.06013} {arXiv:2104.06013 [gr-qc]} \BibitemShut
  {NoStop}%
\bibitem [{\citenamefont {Bonelli}\ \emph {et~al.}(2021)\citenamefont
  {Bonelli}, \citenamefont {Iossa}, \citenamefont {Lichtig},\ and\
  \citenamefont {Tanzini}}]{Bonelli:2021uvf}%
  \BibitemOpen
  \bibfield  {author} {\bibinfo {author} {\bibfnamefont {G.}~\bibnamefont
  {Bonelli}}, \bibinfo {author} {\bibfnamefont {C.}~\bibnamefont {Iossa}},
  \bibinfo {author} {\bibfnamefont {D.~P.}\ \bibnamefont {Lichtig}}, \ and\
  \bibinfo {author} {\bibfnamefont {A.}~\bibnamefont {Tanzini}},\ }\href@noop
  {} {\  (\bibinfo {year} {2021})},\ \Eprint {http://arxiv.org/abs/2105.04483}
  {arXiv:2105.04483 [hep-th]} \BibitemShut {NoStop}%
\bibitem [{\citenamefont {Datta}\ \emph
  {et~al.}(2020{\natexlab{b}})\citenamefont {Datta}, \citenamefont {Phukon},\
  and\ \citenamefont {Bose}}]{Datta:2020gem}%
  \BibitemOpen
  \bibfield  {author} {\bibinfo {author} {\bibfnamefont {S.}~\bibnamefont
  {Datta}}, \bibinfo {author} {\bibfnamefont {K.~S.}\ \bibnamefont {Phukon}}, \
  and\ \bibinfo {author} {\bibfnamefont {S.}~\bibnamefont {Bose}},\ }\href@noop
  {} {\  (\bibinfo {year} {2020}{\natexlab{b}})},\ \Eprint
  {http://arxiv.org/abs/2004.05974} {arXiv:2004.05974 [gr-qc]} \BibitemShut
  {NoStop}%
\bibitem [{\citenamefont {Flanagan}\ and\ \citenamefont
  {Hinderer}(2008)}]{Flanagan:2007ix}%
  \BibitemOpen
  \bibfield  {author} {\bibinfo {author} {\bibfnamefont {E.~E.}\ \bibnamefont
  {Flanagan}}\ and\ \bibinfo {author} {\bibfnamefont {T.}~\bibnamefont
  {Hinderer}},\ }\href {\doibase 10.1103/PhysRevD.77.021502} {\bibfield
  {journal} {\bibinfo  {journal} {Phys. Rev. D}\ }\textbf {\bibinfo {volume}
  {77}},\ \bibinfo {pages} {021502} (\bibinfo {year} {2008})},\ \Eprint
  {http://arxiv.org/abs/0709.1915} {arXiv:0709.1915 [astro-ph]} \BibitemShut
  {NoStop}%
\bibitem [{\citenamefont {Binnington}\ and\ \citenamefont
  {Poisson}(2009)}]{Binnington:2009bb}%
  \BibitemOpen
  \bibfield  {author} {\bibinfo {author} {\bibfnamefont {T.}~\bibnamefont
  {Binnington}}\ and\ \bibinfo {author} {\bibfnamefont {E.}~\bibnamefont
  {Poisson}},\ }\href {\doibase 10.1103/PhysRevD.80.084018} {\bibfield
  {journal} {\bibinfo  {journal} {Phys. Rev.}\ }\textbf {\bibinfo {volume}
  {D80}},\ \bibinfo {pages} {084018} (\bibinfo {year} {2009})},\ \Eprint
  {http://arxiv.org/abs/0906.1366} {arXiv:0906.1366 [gr-qc]} \BibitemShut
  {NoStop}%
\bibitem [{\citenamefont {Landry}\ and\ \citenamefont
  {Poisson}(2014)}]{Landry:2014jka}%
  \BibitemOpen
  \bibfield  {author} {\bibinfo {author} {\bibfnamefont {P.}~\bibnamefont
  {Landry}}\ and\ \bibinfo {author} {\bibfnamefont {E.}~\bibnamefont
  {Poisson}},\ }\href {\doibase 10.1103/PhysRevD.89.124011} {\bibfield
  {journal} {\bibinfo  {journal} {Phys. Rev. D}\ }\textbf {\bibinfo {volume}
  {89}},\ \bibinfo {pages} {124011} (\bibinfo {year} {2014})},\ \Eprint
  {http://arxiv.org/abs/1404.6798} {arXiv:1404.6798 [gr-qc]} \BibitemShut
  {NoStop}%
\bibitem [{\citenamefont {Le~Tiec}\ \emph {et~al.}(2021)\citenamefont
  {Le~Tiec}, \citenamefont {Casals},\ and\ \citenamefont
  {Franzin}}]{LeTiec:2020bos}%
  \BibitemOpen
  \bibfield  {author} {\bibinfo {author} {\bibfnamefont {A.}~\bibnamefont
  {Le~Tiec}}, \bibinfo {author} {\bibfnamefont {M.}~\bibnamefont {Casals}}, \
  and\ \bibinfo {author} {\bibfnamefont {E.}~\bibnamefont {Franzin}},\ }\href
  {\doibase 10.1103/PhysRevD.103.084021} {\bibfield  {journal} {\bibinfo
  {journal} {Phys. Rev. D}\ }\textbf {\bibinfo {volume} {103}},\ \bibinfo
  {pages} {084021} (\bibinfo {year} {2021})},\ \Eprint
  {http://arxiv.org/abs/2010.15795} {arXiv:2010.15795 [gr-qc]} \BibitemShut
  {NoStop}%
\bibitem [{\citenamefont {Chia}(2020)}]{Chia:2020yla}%
  \BibitemOpen
  \bibfield  {author} {\bibinfo {author} {\bibfnamefont {H.~S.}\ \bibnamefont
  {Chia}},\ }\href@noop {} {\  (\bibinfo {year} {2020})},\ \Eprint
  {http://arxiv.org/abs/2010.07300} {arXiv:2010.07300 [gr-qc]} \BibitemShut
  {NoStop}%
\bibitem [{\citenamefont {Charalambous}\ \emph
  {et~al.}(2021{\natexlab{a}})\citenamefont {Charalambous}, \citenamefont
  {Dubovsky},\ and\ \citenamefont {Ivanov}}]{Charalambous:2021kcz}%
  \BibitemOpen
  \bibfield  {author} {\bibinfo {author} {\bibfnamefont {P.}~\bibnamefont
  {Charalambous}}, \bibinfo {author} {\bibfnamefont {S.}~\bibnamefont
  {Dubovsky}}, \ and\ \bibinfo {author} {\bibfnamefont {M.~M.}\ \bibnamefont
  {Ivanov}},\ }\href {\doibase 10.1103/PhysRevLett.127.101101} {\bibfield
  {journal} {\bibinfo  {journal} {Phys. Rev. Lett.}\ }\textbf {\bibinfo
  {volume} {127}},\ \bibinfo {pages} {101101} (\bibinfo {year}
  {2021}{\natexlab{a}})},\ \Eprint {http://arxiv.org/abs/2103.01234}
  {arXiv:2103.01234 [hep-th]} \BibitemShut {NoStop}%
\bibitem [{\citenamefont {Charalambous}\ \emph
  {et~al.}(2021{\natexlab{b}})\citenamefont {Charalambous}, \citenamefont
  {Dubovsky},\ and\ \citenamefont {Ivanov}}]{Charalambous:2021mea}%
  \BibitemOpen
  \bibfield  {author} {\bibinfo {author} {\bibfnamefont {P.}~\bibnamefont
  {Charalambous}}, \bibinfo {author} {\bibfnamefont {S.}~\bibnamefont
  {Dubovsky}}, \ and\ \bibinfo {author} {\bibfnamefont {M.~M.}\ \bibnamefont
  {Ivanov}},\ }\href {\doibase 10.1007/JHEP05(2021)038} {\bibfield  {journal}
  {\bibinfo  {journal} {JHEP}\ }\textbf {\bibinfo {volume} {05}},\ \bibinfo
  {pages} {038} (\bibinfo {year} {2021}{\natexlab{b}})},\ \Eprint
  {http://arxiv.org/abs/2102.08917} {arXiv:2102.08917 [hep-th]} \BibitemShut
  {NoStop}%
\bibitem [{\citenamefont {Hui}\ \emph {et~al.}(2022{\natexlab{a}})\citenamefont
  {Hui}, \citenamefont {Joyce}, \citenamefont {Penco}, \citenamefont
  {Santoni},\ and\ \citenamefont {Solomon}}]{Hui:2021vcv}%
  \BibitemOpen
  \bibfield  {author} {\bibinfo {author} {\bibfnamefont {L.}~\bibnamefont
  {Hui}}, \bibinfo {author} {\bibfnamefont {A.}~\bibnamefont {Joyce}}, \bibinfo
  {author} {\bibfnamefont {R.}~\bibnamefont {Penco}}, \bibinfo {author}
  {\bibfnamefont {L.}~\bibnamefont {Santoni}}, \ and\ \bibinfo {author}
  {\bibfnamefont {A.~R.}\ \bibnamefont {Solomon}},\ }\href {\doibase
  10.1088/1475-7516/2022/01/032} {\bibfield  {journal} {\bibinfo  {journal}
  {JCAP}\ }\textbf {\bibinfo {volume} {01}},\ \bibinfo {pages} {032} (\bibinfo
  {year} {2022}{\natexlab{a}})},\ \Eprint {http://arxiv.org/abs/2105.01069}
  {arXiv:2105.01069 [hep-th]} \BibitemShut {NoStop}%
\bibitem [{\citenamefont {Hui}\ \emph {et~al.}(2022{\natexlab{b}})\citenamefont
  {Hui}, \citenamefont {Joyce}, \citenamefont {Penco}, \citenamefont
  {Santoni},\ and\ \citenamefont {Solomon}}]{Hui:2022vbh}%
  \BibitemOpen
  \bibfield  {author} {\bibinfo {author} {\bibfnamefont {L.}~\bibnamefont
  {Hui}}, \bibinfo {author} {\bibfnamefont {A.}~\bibnamefont {Joyce}}, \bibinfo
  {author} {\bibfnamefont {R.}~\bibnamefont {Penco}}, \bibinfo {author}
  {\bibfnamefont {L.}~\bibnamefont {Santoni}}, \ and\ \bibinfo {author}
  {\bibfnamefont {A.~R.}\ \bibnamefont {Solomon}},\ }\href@noop {} {\
  (\bibinfo {year} {2022}{\natexlab{b}})},\ \Eprint
  {http://arxiv.org/abs/2203.08832} {arXiv:2203.08832 [hep-th]} \BibitemShut
  {NoStop}%
\bibitem [{\citenamefont {Ben~Achour}\ \emph {et~al.}(2022)\citenamefont
  {Ben~Achour}, \citenamefont {Livine}, \citenamefont {Mukohyama},\ and\
  \citenamefont {Uzan}}]{BenAchour:2022uqo}%
  \BibitemOpen
  \bibfield  {author} {\bibinfo {author} {\bibfnamefont {J.}~\bibnamefont
  {Ben~Achour}}, \bibinfo {author} {\bibfnamefont {E.~R.}\ \bibnamefont
  {Livine}}, \bibinfo {author} {\bibfnamefont {S.}~\bibnamefont {Mukohyama}}, \
  and\ \bibinfo {author} {\bibfnamefont {J.-P.}\ \bibnamefont {Uzan}},\ }\href
  {\doibase 10.1007/JHEP07(2022)112} {\bibfield  {journal} {\bibinfo  {journal}
  {JHEP}\ }\textbf {\bibinfo {volume} {07}},\ \bibinfo {pages} {112} (\bibinfo
  {year} {2022})},\ \Eprint {http://arxiv.org/abs/2202.12828} {arXiv:2202.12828
  [gr-qc]} \BibitemShut {NoStop}%
\bibitem [{\citenamefont {Hinderer}(2008)}]{Hinderer:2007mb}%
  \BibitemOpen
  \bibfield  {author} {\bibinfo {author} {\bibfnamefont {T.}~\bibnamefont
  {Hinderer}},\ }\href {\doibase 10.1086/533487} {\bibfield  {journal}
  {\bibinfo  {journal} {Astrophys. J.}\ }\textbf {\bibinfo {volume} {677}},\
  \bibinfo {pages} {1216} (\bibinfo {year} {2008})},\ \Eprint
  {http://arxiv.org/abs/0711.2420} {arXiv:0711.2420 [astro-ph]} \BibitemShut
  {NoStop}%
\bibitem [{\citenamefont {Hinderer}\ \emph {et~al.}(2010)\citenamefont
  {Hinderer}, \citenamefont {Lackey}, \citenamefont {Lang},\ and\ \citenamefont
  {Read}}]{Hinderer:2009ca}%
  \BibitemOpen
  \bibfield  {author} {\bibinfo {author} {\bibfnamefont {T.}~\bibnamefont
  {Hinderer}}, \bibinfo {author} {\bibfnamefont {B.~D.}\ \bibnamefont
  {Lackey}}, \bibinfo {author} {\bibfnamefont {R.~N.}\ \bibnamefont {Lang}}, \
  and\ \bibinfo {author} {\bibfnamefont {J.~S.}\ \bibnamefont {Read}},\ }\href
  {\doibase 10.1103/PhysRevD.81.123016} {\bibfield  {journal} {\bibinfo
  {journal} {Phys. Rev. D}\ }\textbf {\bibinfo {volume} {81}},\ \bibinfo
  {pages} {123016} (\bibinfo {year} {2010})},\ \Eprint
  {http://arxiv.org/abs/0911.3535} {arXiv:0911.3535 [astro-ph.HE]} \BibitemShut
  {NoStop}%
\bibitem [{\citenamefont {Abbott}\ \emph
  {et~al.}(2017{\natexlab{b}})\citenamefont {Abbott} \emph
  {et~al.}}]{TheLIGOScientific:2017qsa}%
  \BibitemOpen
  \bibfield  {author} {\bibinfo {author} {\bibfnamefont {B.}~\bibnamefont
  {Abbott}} \emph {et~al.} (\bibinfo {collaboration} {LIGO Scientific,
  Virgo}),\ }\href {\doibase 10.1103/PhysRevLett.119.161101} {\bibfield
  {journal} {\bibinfo  {journal} {Phys. Rev. Lett.}\ }\textbf {\bibinfo
  {volume} {119}},\ \bibinfo {pages} {161101} (\bibinfo {year}
  {2017}{\natexlab{b}})},\ \Eprint {http://arxiv.org/abs/1710.05832}
  {arXiv:1710.05832 [gr-qc]} \BibitemShut {NoStop}%
\bibitem [{\citenamefont {Abbott}\ \emph {et~al.}(2018)\citenamefont {Abbott}
  \emph {et~al.}}]{Abbott:2018exr}%
  \BibitemOpen
  \bibfield  {author} {\bibinfo {author} {\bibfnamefont {B.~P.}\ \bibnamefont
  {Abbott}} \emph {et~al.} (\bibinfo {collaboration} {LIGO Scientific,
  Virgo}),\ }\href {\doibase 10.1103/PhysRevLett.121.161101} {\bibfield
  {journal} {\bibinfo  {journal} {Phys. Rev. Lett.}\ }\textbf {\bibinfo
  {volume} {121}},\ \bibinfo {pages} {161101} (\bibinfo {year} {2018})},\
  \Eprint {http://arxiv.org/abs/1805.11581} {arXiv:1805.11581 [gr-qc]}
  \BibitemShut {NoStop}%
\bibitem [{\citenamefont {Char}\ and\ \citenamefont
  {Datta}(2018)}]{Char:2018grw}%
  \BibitemOpen
  \bibfield  {author} {\bibinfo {author} {\bibfnamefont {P.}~\bibnamefont
  {Char}}\ and\ \bibinfo {author} {\bibfnamefont {S.}~\bibnamefont {Datta}},\
  }\href {\doibase 10.1103/PhysRevD.98.084010} {\bibfield  {journal} {\bibinfo
  {journal} {Phys. Rev. D}\ }\textbf {\bibinfo {volume} {98}},\ \bibinfo
  {pages} {084010} (\bibinfo {year} {2018})},\ \Eprint
  {http://arxiv.org/abs/1806.10986} {arXiv:1806.10986 [gr-qc]} \BibitemShut
  {NoStop}%
\bibitem [{\citenamefont {Datta}\ and\ \citenamefont
  {Char}(2020)}]{Datta:2019ueq}%
  \BibitemOpen
  \bibfield  {author} {\bibinfo {author} {\bibfnamefont {S.}~\bibnamefont
  {Datta}}\ and\ \bibinfo {author} {\bibfnamefont {P.}~\bibnamefont {Char}},\
  }\href {\doibase 10.1103/PhysRevD.101.064016} {\bibfield  {journal} {\bibinfo
   {journal} {Phys. Rev. D}\ }\textbf {\bibinfo {volume} {101}},\ \bibinfo
  {pages} {064016} (\bibinfo {year} {2020})},\ \Eprint
  {http://arxiv.org/abs/1908.04235} {arXiv:1908.04235 [gr-qc]} \BibitemShut
  {NoStop}%
\bibitem [{\citenamefont {Raposo}\ \emph {et~al.}(2019)\citenamefont {Raposo},
  \citenamefont {Pani}, \citenamefont {Bezares}, \citenamefont {Palenzuela},\
  and\ \citenamefont {Cardoso}}]{Raposo:2018rjn}%
  \BibitemOpen
  \bibfield  {author} {\bibinfo {author} {\bibfnamefont {G.}~\bibnamefont
  {Raposo}}, \bibinfo {author} {\bibfnamefont {P.}~\bibnamefont {Pani}},
  \bibinfo {author} {\bibfnamefont {M.}~\bibnamefont {Bezares}}, \bibinfo
  {author} {\bibfnamefont {C.}~\bibnamefont {Palenzuela}}, \ and\ \bibinfo
  {author} {\bibfnamefont {V.}~\bibnamefont {Cardoso}},\ }\href {\doibase
  10.1103/PhysRevD.99.104072} {\bibfield  {journal} {\bibinfo  {journal} {Phys.
  Rev. D}\ }\textbf {\bibinfo {volume} {99}},\ \bibinfo {pages} {104072}
  (\bibinfo {year} {2019})},\ \Eprint {http://arxiv.org/abs/1811.07917}
  {arXiv:1811.07917 [gr-qc]} \BibitemShut {NoStop}%
\bibitem [{\citenamefont {Biswas}\ and\ \citenamefont
  {Bose}(2019)}]{Biswas:2019gkw}%
  \BibitemOpen
  \bibfield  {author} {\bibinfo {author} {\bibfnamefont {B.}~\bibnamefont
  {Biswas}}\ and\ \bibinfo {author} {\bibfnamefont {S.}~\bibnamefont {Bose}},\
  }\href {\doibase 10.1103/PhysRevD.99.104002} {\bibfield  {journal} {\bibinfo
  {journal} {Phys. Rev. D}\ }\textbf {\bibinfo {volume} {99}},\ \bibinfo
  {pages} {104002} (\bibinfo {year} {2019})},\ \Eprint
  {http://arxiv.org/abs/1903.04956} {arXiv:1903.04956 [gr-qc]} \BibitemShut
  {NoStop}%
\bibitem [{\citenamefont {Baiotti}(2019)}]{Baiotti:2019sew}%
  \BibitemOpen
  \bibfield  {author} {\bibinfo {author} {\bibfnamefont {L.}~\bibnamefont
  {Baiotti}},\ }\href {\doibase 10.1016/j.ppnp.2019.103714} {\bibfield
  {journal} {\bibinfo  {journal} {Prog. Part. Nucl. Phys.}\ }\textbf {\bibinfo
  {volume} {109}},\ \bibinfo {pages} {103714} (\bibinfo {year} {2019})},\
  \Eprint {http://arxiv.org/abs/1907.08534} {arXiv:1907.08534 [astro-ph.HE]}
  \BibitemShut {NoStop}%
\bibitem [{\citenamefont {Dietrich}\ \emph {et~al.}(2021)\citenamefont
  {Dietrich}, \citenamefont {Hinderer},\ and\ \citenamefont
  {Samajdar}}]{Dietrich:2020eud}%
  \BibitemOpen
  \bibfield  {author} {\bibinfo {author} {\bibfnamefont {T.}~\bibnamefont
  {Dietrich}}, \bibinfo {author} {\bibfnamefont {T.}~\bibnamefont {Hinderer}},
  \ and\ \bibinfo {author} {\bibfnamefont {A.}~\bibnamefont {Samajdar}},\
  }\href {\doibase 10.1007/s10714-020-02751-6} {\bibfield  {journal} {\bibinfo
  {journal} {Gen. Rel. Grav.}\ }\textbf {\bibinfo {volume} {53}},\ \bibinfo
  {pages} {27} (\bibinfo {year} {2021})},\ \Eprint
  {http://arxiv.org/abs/2004.02527} {arXiv:2004.02527 [gr-qc]} \BibitemShut
  {NoStop}%
\bibitem [{\citenamefont {Addazi}\ \emph {et~al.}(2019)\citenamefont {Addazi},
  \citenamefont {Marciano},\ and\ \citenamefont {Yunes}}]{Addazi:2018uhd}%
  \BibitemOpen
  \bibfield  {author} {\bibinfo {author} {\bibfnamefont {A.}~\bibnamefont
  {Addazi}}, \bibinfo {author} {\bibfnamefont {A.}~\bibnamefont {Marciano}}, \
  and\ \bibinfo {author} {\bibfnamefont {N.}~\bibnamefont {Yunes}},\ }\href
  {\doibase 10.1103/PhysRevLett.122.081301} {\bibfield  {journal} {\bibinfo
  {journal} {Phys. Rev. Lett.}\ }\textbf {\bibinfo {volume} {122}},\ \bibinfo
  {pages} {081301} (\bibinfo {year} {2019})},\ \Eprint
  {http://arxiv.org/abs/1810.10417} {arXiv:1810.10417 [gr-qc]} \BibitemShut
  {NoStop}%
\bibitem [{\citenamefont {Amaro-Seoane}\ \emph {et~al.}(2017)\citenamefont
  {Amaro-Seoane} \emph {et~al.}}]{Audley:2017drz}%
  \BibitemOpen
  \bibfield  {author} {\bibinfo {author} {\bibfnamefont {P.}~\bibnamefont
  {Amaro-Seoane}} \emph {et~al.} (\bibinfo {collaboration} {LISA}),\
  }\href@noop {} {\  (\bibinfo {year} {2017})},\ \Eprint
  {http://arxiv.org/abs/1702.00786} {arXiv:1702.00786 [astro-ph.IM]}
  \BibitemShut {NoStop}%
\bibitem [{\citenamefont {Gair}\ \emph {et~al.}(2004)\citenamefont {Gair},
  \citenamefont {Barack}, \citenamefont {Creighton}, \citenamefont {Cutler},
  \citenamefont {Larson}, \citenamefont {Phinney},\ and\ \citenamefont
  {Vallisneri}}]{Gair:2004iv}%
  \BibitemOpen
  \bibfield  {author} {\bibinfo {author} {\bibfnamefont {J.~R.}\ \bibnamefont
  {Gair}}, \bibinfo {author} {\bibfnamefont {L.}~\bibnamefont {Barack}},
  \bibinfo {author} {\bibfnamefont {T.}~\bibnamefont {Creighton}}, \bibinfo
  {author} {\bibfnamefont {C.}~\bibnamefont {Cutler}}, \bibinfo {author}
  {\bibfnamefont {S.~L.}\ \bibnamefont {Larson}}, \bibinfo {author}
  {\bibfnamefont {E.~S.}\ \bibnamefont {Phinney}}, \ and\ \bibinfo {author}
  {\bibfnamefont {M.}~\bibnamefont {Vallisneri}},\ }\href {\doibase
  10.1088/0264-9381/21/20/003} {\bibfield  {journal} {\bibinfo  {journal}
  {Class. Quant. Grav.}\ }\textbf {\bibinfo {volume} {21}},\ \bibinfo {pages}
  {S1595} (\bibinfo {year} {2004})},\ \Eprint
  {http://arxiv.org/abs/gr-qc/0405137} {arXiv:gr-qc/0405137} \BibitemShut
  {NoStop}%
\bibitem [{\citenamefont {Gair}\ and\ \citenamefont
  {Porter}(2013)}]{Gair:2012vi}%
  \BibitemOpen
  \bibfield  {author} {\bibinfo {author} {\bibfnamefont {J.~R.}\ \bibnamefont
  {Gair}}\ and\ \bibinfo {author} {\bibfnamefont {E.~K.}\ \bibnamefont
  {Porter}},\ }\href@noop {} {\bibfield  {journal} {\bibinfo  {journal} {ASP
  Conf. Ser.}\ }\textbf {\bibinfo {volume} {467}},\ \bibinfo {pages} {173}
  (\bibinfo {year} {2013})},\ \Eprint {http://arxiv.org/abs/1210.8066}
  {arXiv:1210.8066 [gr-qc]} \BibitemShut {NoStop}%
\bibitem [{\citenamefont {Amaro-Seoane}(2019)}]{Amaro-Seoane:2019umn}%
  \BibitemOpen
  \bibfield  {author} {\bibinfo {author} {\bibfnamefont {P.}~\bibnamefont
  {Amaro-Seoane}},\ }\href {\doibase 10.1103/PhysRevD.99.123025} {\bibfield
  {journal} {\bibinfo  {journal} {Phys. Rev. D}\ }\textbf {\bibinfo {volume}
  {99}},\ \bibinfo {pages} {123025} (\bibinfo {year} {2019})},\ \Eprint
  {http://arxiv.org/abs/1903.10871} {arXiv:1903.10871 [astro-ph.GA]}
  \BibitemShut {NoStop}%
\bibitem [{\citenamefont {Pani}\ and\ \citenamefont
  {Maselli}(2019)}]{Pani:2019cyc}%
  \BibitemOpen
  \bibfield  {author} {\bibinfo {author} {\bibfnamefont {P.}~\bibnamefont
  {Pani}}\ and\ \bibinfo {author} {\bibfnamefont {A.}~\bibnamefont {Maselli}},\
  }\href {\doibase 10.1142/S0218271819440012} {\bibfield  {journal} {\bibinfo
  {journal} {Int. J. Mod. Phys. D}\ }\textbf {\bibinfo {volume} {28}},\
  \bibinfo {pages} {1944001} (\bibinfo {year} {2019})},\ \Eprint
  {http://arxiv.org/abs/1905.03947} {arXiv:1905.03947 [gr-qc]} \BibitemShut
  {NoStop}%
\bibitem [{\citenamefont {Flanagan}\ and\ \citenamefont
  {Hughes}(1998)}]{Flanagan:1997kp}%
  \BibitemOpen
  \bibfield  {author} {\bibinfo {author} {\bibfnamefont {E.~E.}\ \bibnamefont
  {Flanagan}}\ and\ \bibinfo {author} {\bibfnamefont {S.~A.}\ \bibnamefont
  {Hughes}},\ }\href {\doibase 10.1103/PhysRevD.57.4566} {\bibfield  {journal}
  {\bibinfo  {journal} {Phys. Rev. D}\ }\textbf {\bibinfo {volume} {57}},\
  \bibinfo {pages} {4566} (\bibinfo {year} {1998})},\ \Eprint
  {http://arxiv.org/abs/gr-qc/9710129} {arXiv:gr-qc/9710129} \BibitemShut
  {NoStop}%
\bibitem [{\citenamefont {Lindblom}\ \emph {et~al.}(2008)\citenamefont
  {Lindblom}, \citenamefont {Owen},\ and\ \citenamefont
  {Brown}}]{Lindblom:2008cm}%
  \BibitemOpen
  \bibfield  {author} {\bibinfo {author} {\bibfnamefont {L.}~\bibnamefont
  {Lindblom}}, \bibinfo {author} {\bibfnamefont {B.~J.}\ \bibnamefont {Owen}},
  \ and\ \bibinfo {author} {\bibfnamefont {D.~A.}\ \bibnamefont {Brown}},\
  }\href {\doibase 10.1103/PhysRevD.78.124020} {\bibfield  {journal} {\bibinfo
  {journal} {Phys. Rev. D}\ }\textbf {\bibinfo {volume} {78}},\ \bibinfo
  {pages} {124020} (\bibinfo {year} {2008})},\ \Eprint
  {http://arxiv.org/abs/0809.3844} {arXiv:0809.3844 [gr-qc]} \BibitemShut
  {NoStop}%
\bibitem [{\citenamefont {Amaro-Seoane}\ \emph {et~al.}(2007)\citenamefont
  {Amaro-Seoane}, \citenamefont {Gair}, \citenamefont {Freitag}, \citenamefont
  {Coleman~Miller}, \citenamefont {Mandel}, \citenamefont {Cutler},\ and\
  \citenamefont {Babak}}]{Amaro-Seoane:2007osp}%
  \BibitemOpen
  \bibfield  {author} {\bibinfo {author} {\bibfnamefont {P.}~\bibnamefont
  {Amaro-Seoane}}, \bibinfo {author} {\bibfnamefont {J.~R.}\ \bibnamefont
  {Gair}}, \bibinfo {author} {\bibfnamefont {M.}~\bibnamefont {Freitag}},
  \bibinfo {author} {\bibfnamefont {M.}~\bibnamefont {Coleman~Miller}},
  \bibinfo {author} {\bibfnamefont {I.}~\bibnamefont {Mandel}}, \bibinfo
  {author} {\bibfnamefont {C.~J.}\ \bibnamefont {Cutler}}, \ and\ \bibinfo
  {author} {\bibfnamefont {S.}~\bibnamefont {Babak}},\ }\href {\doibase
  10.1088/0264-9381/24/17/R01} {\bibfield  {journal} {\bibinfo  {journal}
  {Class. Quant. Grav.}\ }\textbf {\bibinfo {volume} {24}},\ \bibinfo {pages}
  {R113} (\bibinfo {year} {2007})},\ \Eprint
  {http://arxiv.org/abs/astro-ph/0703495} {arXiv:astro-ph/0703495} \BibitemShut
  {NoStop}%
\bibitem [{\citenamefont {Porter}\ and\ \citenamefont
  {Sesana}(2010)}]{Porter:2010mb}%
  \BibitemOpen
  \bibfield  {author} {\bibinfo {author} {\bibfnamefont {E.~K.}\ \bibnamefont
  {Porter}}\ and\ \bibinfo {author} {\bibfnamefont {A.}~\bibnamefont
  {Sesana}},\ }\href@noop {} {\  (\bibinfo {year} {2010})},\ \Eprint
  {http://arxiv.org/abs/1005.5296} {arXiv:1005.5296 [gr-qc]} \BibitemShut
  {NoStop}%
\bibitem [{\citenamefont {Amaro-Seoane}\ \emph {et~al.}(2011)\citenamefont
  {Amaro-Seoane}, \citenamefont {Schutz},\ and\ \citenamefont
  {Thornburg}}]{Amaro-Seoane:2011fgy}%
  \BibitemOpen
  \bibfield  {author} {\bibinfo {author} {\bibfnamefont {P.}~\bibnamefont
  {Amaro-Seoane}}, \bibinfo {author} {\bibfnamefont {B.~F.}\ \bibnamefont
  {Schutz}}, \ and\ \bibinfo {author} {\bibfnamefont {J.}~\bibnamefont
  {Thornburg}},\ }\href@noop {} {\  (\bibinfo {year} {2011})},\ \Eprint
  {http://arxiv.org/abs/1102.3647} {arXiv:1102.3647 [astro-ph.CO]} \BibitemShut
  {NoStop}%
\bibitem [{\citenamefont {Katz}\ \emph {et~al.}(2021)\citenamefont {Katz},
  \citenamefont {Chua}, \citenamefont {Speri}, \citenamefont {Warburton},\ and\
  \citenamefont {Hughes}}]{Katz:2021yft}%
  \BibitemOpen
  \bibfield  {author} {\bibinfo {author} {\bibfnamefont {M.~L.}\ \bibnamefont
  {Katz}}, \bibinfo {author} {\bibfnamefont {A.~J.~K.}\ \bibnamefont {Chua}},
  \bibinfo {author} {\bibfnamefont {L.}~\bibnamefont {Speri}}, \bibinfo
  {author} {\bibfnamefont {N.}~\bibnamefont {Warburton}}, \ and\ \bibinfo
  {author} {\bibfnamefont {S.~A.}\ \bibnamefont {Hughes}},\ }\href {\doibase
  10.1103/PhysRevD.104.064047} {\bibfield  {journal} {\bibinfo  {journal}
  {Phys. Rev. D}\ }\textbf {\bibinfo {volume} {104}},\ \bibinfo {pages}
  {064047} (\bibinfo {year} {2021})},\ \Eprint
  {http://arxiv.org/abs/2104.04582} {arXiv:2104.04582 [gr-qc]} \BibitemShut
  {NoStop}%
\bibitem [{\citenamefont {Robson}\ \emph {et~al.}(2019)\citenamefont {Robson},
  \citenamefont {Cornish},\ and\ \citenamefont {Liu}}]{Robson:2018ifk}%
  \BibitemOpen
  \bibfield  {author} {\bibinfo {author} {\bibfnamefont {T.}~\bibnamefont
  {Robson}}, \bibinfo {author} {\bibfnamefont {N.~J.}\ \bibnamefont {Cornish}},
  \ and\ \bibinfo {author} {\bibfnamefont {C.}~\bibnamefont {Liu}},\ }\href
  {\doibase 10.1088/1361-6382/ab1101} {\bibfield  {journal} {\bibinfo
  {journal} {Class. Quant. Grav.}\ }\textbf {\bibinfo {volume} {36}},\ \bibinfo
  {pages} {105011} (\bibinfo {year} {2019})},\ \Eprint
  {http://arxiv.org/abs/1803.01944} {arXiv:1803.01944 [astro-ph.HE]}
  \BibitemShut {NoStop}%
\bibitem [{\citenamefont {Poisson}(1996)}]{Poisson:1996tc}%
  \BibitemOpen
  \bibfield  {author} {\bibinfo {author} {\bibfnamefont {E.}~\bibnamefont
  {Poisson}},\ }\href {\doibase 10.1103/PhysRevD.54.5939} {\bibfield  {journal}
  {\bibinfo  {journal} {Phys. Rev. D}\ }\textbf {\bibinfo {volume} {54}},\
  \bibinfo {pages} {5939} (\bibinfo {year} {1996})},\ \Eprint
  {http://arxiv.org/abs/gr-qc/9606024} {arXiv:gr-qc/9606024} \BibitemShut
  {NoStop}%
\bibitem [{\citenamefont {Barack}\ and\ \citenamefont
  {Cutler}(2004)}]{Barack:2003fp}%
  \BibitemOpen
  \bibfield  {author} {\bibinfo {author} {\bibfnamefont {L.}~\bibnamefont
  {Barack}}\ and\ \bibinfo {author} {\bibfnamefont {C.}~\bibnamefont
  {Cutler}},\ }\href {\doibase 10.1103/PhysRevD.69.082005} {\bibfield
  {journal} {\bibinfo  {journal} {Phys. Rev. D}\ }\textbf {\bibinfo {volume}
  {69}},\ \bibinfo {pages} {082005} (\bibinfo {year} {2004})},\ \Eprint
  {http://arxiv.org/abs/gr-qc/0310125} {arXiv:gr-qc/0310125} \BibitemShut
  {NoStop}%
\bibitem [{\citenamefont {Kim}\ and\ \citenamefont {Shim}(2021)}]{Kim:2020dif}%
  \BibitemOpen
  \bibfield  {author} {\bibinfo {author} {\bibfnamefont {J.-W.}\ \bibnamefont
  {Kim}}\ and\ \bibinfo {author} {\bibfnamefont {M.}~\bibnamefont {Shim}},\
  }\href {\doibase 10.1103/PhysRevD.104.046022} {\bibfield  {journal} {\bibinfo
   {journal} {Phys. Rev. D}\ }\textbf {\bibinfo {volume} {104}},\ \bibinfo
  {pages} {046022} (\bibinfo {year} {2021})},\ \Eprint
  {http://arxiv.org/abs/2011.03337} {arXiv:2011.03337 [hep-th]} \BibitemShut
  {NoStop}%
\bibitem [{\citenamefont {Yunes}\ \emph
  {et~al.}(2011{\natexlab{a}})\citenamefont {Yunes}, \citenamefont
  {Coleman~Miller},\ and\ \citenamefont {Thornburg}}]{Yunes:2010sm}%
  \BibitemOpen
  \bibfield  {author} {\bibinfo {author} {\bibfnamefont {N.}~\bibnamefont
  {Yunes}}, \bibinfo {author} {\bibfnamefont {M.}~\bibnamefont
  {Coleman~Miller}}, \ and\ \bibinfo {author} {\bibfnamefont {J.}~\bibnamefont
  {Thornburg}},\ }\href {\doibase 10.1103/PhysRevD.83.044030} {\bibfield
  {journal} {\bibinfo  {journal} {Phys. Rev. D}\ }\textbf {\bibinfo {volume}
  {83}},\ \bibinfo {pages} {044030} (\bibinfo {year} {2011}{\natexlab{a}})},\
  \Eprint {http://arxiv.org/abs/1010.1721} {arXiv:1010.1721 [astro-ph.GA]}
  \BibitemShut {NoStop}%
\bibitem [{\citenamefont {Yunes}\ \emph
  {et~al.}(2011{\natexlab{b}})\citenamefont {Yunes}, \citenamefont {Kocsis},
  \citenamefont {Loeb},\ and\ \citenamefont {Haiman}}]{Yunes:2011ws}%
  \BibitemOpen
  \bibfield  {author} {\bibinfo {author} {\bibfnamefont {N.}~\bibnamefont
  {Yunes}}, \bibinfo {author} {\bibfnamefont {B.}~\bibnamefont {Kocsis}},
  \bibinfo {author} {\bibfnamefont {A.}~\bibnamefont {Loeb}}, \ and\ \bibinfo
  {author} {\bibfnamefont {Z.}~\bibnamefont {Haiman}},\ }\href {\doibase
  10.1103/PhysRevLett.107.171103} {\bibfield  {journal} {\bibinfo  {journal}
  {Phys. Rev. Lett.}\ }\textbf {\bibinfo {volume} {107}},\ \bibinfo {pages}
  {171103} (\bibinfo {year} {2011}{\natexlab{b}})},\ \Eprint
  {http://arxiv.org/abs/1103.4609} {arXiv:1103.4609 [astro-ph.CO]} \BibitemShut
  {NoStop}%
\bibitem [{\citenamefont {Kocsis}\ \emph {et~al.}(2011)\citenamefont {Kocsis},
  \citenamefont {Yunes},\ and\ \citenamefont {Loeb}}]{Kocsis:2011dr}%
  \BibitemOpen
  \bibfield  {author} {\bibinfo {author} {\bibfnamefont {B.}~\bibnamefont
  {Kocsis}}, \bibinfo {author} {\bibfnamefont {N.}~\bibnamefont {Yunes}}, \
  and\ \bibinfo {author} {\bibfnamefont {A.}~\bibnamefont {Loeb}},\ }\href
  {\doibase 10.1103/PhysRevD.86.049907} {\bibfield  {journal} {\bibinfo
  {journal} {Phys. Rev. D}\ }\textbf {\bibinfo {volume} {84}},\ \bibinfo
  {pages} {024032} (\bibinfo {year} {2011})},\ \Eprint
  {http://arxiv.org/abs/1104.2322} {arXiv:1104.2322 [astro-ph.GA]} \BibitemShut
  {NoStop}%
\bibitem [{\citenamefont {Barausse}\ \emph {et~al.}(2015)\citenamefont
  {Barausse}, \citenamefont {Cardoso},\ and\ \citenamefont
  {Pani}}]{Barausse:2014pra}%
  \BibitemOpen
  \bibfield  {author} {\bibinfo {author} {\bibfnamefont {E.}~\bibnamefont
  {Barausse}}, \bibinfo {author} {\bibfnamefont {V.}~\bibnamefont {Cardoso}}, \
  and\ \bibinfo {author} {\bibfnamefont {P.}~\bibnamefont {Pani}},\ }\href
  {\doibase 10.1088/1742-6596/610/1/012044} {\bibfield  {journal} {\bibinfo
  {journal} {J. Phys. Conf. Ser.}\ }\textbf {\bibinfo {volume} {610}},\
  \bibinfo {pages} {012044} (\bibinfo {year} {2015})},\ \Eprint
  {http://arxiv.org/abs/1404.7140} {arXiv:1404.7140 [astro-ph.CO]} \BibitemShut
  {NoStop}%
\bibitem [{\citenamefont {Cardoso}\ and\ \citenamefont
  {Duque}(2020)}]{Cardoso:2019upw}%
  \BibitemOpen
  \bibfield  {author} {\bibinfo {author} {\bibfnamefont {V.}~\bibnamefont
  {Cardoso}}\ and\ \bibinfo {author} {\bibfnamefont {F.}~\bibnamefont
  {Duque}},\ }\href {\doibase 10.1103/PhysRevD.101.064028} {\bibfield
  {journal} {\bibinfo  {journal} {Phys. Rev. D}\ }\textbf {\bibinfo {volume}
  {101}},\ \bibinfo {pages} {064028} (\bibinfo {year} {2020})},\ \Eprint
  {http://arxiv.org/abs/1912.07616} {arXiv:1912.07616 [gr-qc]} \BibitemShut
  {NoStop}%
\bibitem [{\citenamefont {Addazi}\ \emph {et~al.}(2020)\citenamefont {Addazi},
  \citenamefont {Marcian\`o},\ and\ \citenamefont {Yunes}}]{Addazi:2019bjz}%
  \BibitemOpen
  \bibfield  {author} {\bibinfo {author} {\bibfnamefont {A.}~\bibnamefont
  {Addazi}}, \bibinfo {author} {\bibfnamefont {A.}~\bibnamefont {Marcian\`o}},
  \ and\ \bibinfo {author} {\bibfnamefont {N.}~\bibnamefont {Yunes}},\ }\href
  {\doibase 10.1140/epjc/s10052-019-7575-9} {\bibfield  {journal} {\bibinfo
  {journal} {Eur. Phys. J. C}\ }\textbf {\bibinfo {volume} {80}},\ \bibinfo
  {pages} {36} (\bibinfo {year} {2020})},\ \Eprint
  {http://arxiv.org/abs/1905.08734} {arXiv:1905.08734 [gr-qc]} \BibitemShut
  {NoStop}%
\bibitem [{\citenamefont {Allen}\ \emph {et~al.}(2012)\citenamefont {Allen},
  \citenamefont {Anderson}, \citenamefont {Brady}, \citenamefont {Brown},\ and\
  \citenamefont {Creighton}}]{Allen:2005fk}%
  \BibitemOpen
  \bibfield  {author} {\bibinfo {author} {\bibfnamefont {B.}~\bibnamefont
  {Allen}}, \bibinfo {author} {\bibfnamefont {W.~G.}\ \bibnamefont {Anderson}},
  \bibinfo {author} {\bibfnamefont {P.~R.}\ \bibnamefont {Brady}}, \bibinfo
  {author} {\bibfnamefont {D.~A.}\ \bibnamefont {Brown}}, \ and\ \bibinfo
  {author} {\bibfnamefont {J.~D.~E.}\ \bibnamefont {Creighton}},\ }\href
  {\doibase 10.1103/PhysRevD.85.122006} {\bibfield  {journal} {\bibinfo
  {journal} {Phys. Rev. D}\ }\textbf {\bibinfo {volume} {85}},\ \bibinfo
  {pages} {122006} (\bibinfo {year} {2012})},\ \Eprint
  {http://arxiv.org/abs/gr-qc/0509116} {arXiv:gr-qc/0509116} \BibitemShut
  {NoStop}%
\bibitem [{\citenamefont {Chatziioannou}\ \emph {et~al.}(2017)\citenamefont
  {Chatziioannou}, \citenamefont {Klein}, \citenamefont {Yunes},\ and\
  \citenamefont {Cornish}}]{Chatziioannou:2017tdw}%
  \BibitemOpen
  \bibfield  {author} {\bibinfo {author} {\bibfnamefont {K.}~\bibnamefont
  {Chatziioannou}}, \bibinfo {author} {\bibfnamefont {A.}~\bibnamefont
  {Klein}}, \bibinfo {author} {\bibfnamefont {N.}~\bibnamefont {Yunes}}, \ and\
  \bibinfo {author} {\bibfnamefont {N.}~\bibnamefont {Cornish}},\ }\href
  {\doibase 10.1103/PhysRevD.95.104004} {\bibfield  {journal} {\bibinfo
  {journal} {Phys. Rev. D}\ }\textbf {\bibinfo {volume} {95}},\ \bibinfo
  {pages} {104004} (\bibinfo {year} {2017})},\ \Eprint
  {http://arxiv.org/abs/1703.03967} {arXiv:1703.03967 [gr-qc]} \BibitemShut
  {NoStop}%
\end{thebibliography}%

\end{document}